\documentclass[aps,prl,superscriptaddress,floatfix,showpacs] {revtex4}
\usepackage{amsmath,amssymb,epsfig,pstricks,bm,graphics,alltt}

\begin{document}

\title{Polaron model of pseudogap state in quasi-one-dimentional systems}

\author{Yu.S. Orlov}
\affiliation{L.V. Kirensky Institute of Physics, Siberian Branch, Russian Academy of Sciences, Akademgorodok  50/38, Krasnoyarsk, 660036, Russia}%
\affiliation{Siberian Federal University, Krasnoyarsk, 660041, Russia}%
\author{V.A. Dudnikov}
\affiliation{L.V. Kirensky Institute of Physics, Siberian Branch, Russian Academy of Sciences, Akademgorodok  50/38, Krasnoyarsk, 660036, Russia}%


\begin{abstract}
A brief overview of the basic concepts and problems of the physics of quasi-one-dimensional compounds is done.  Consistent theoretical description of the so-called pseudogap state remains the main problem. A simplified model of the pseudogap state, based on the picture of small polarons formation in the framework of cluster perturbation theory is considered.
\end{abstract}


\maketitle

\section{\label{sec:Intr}Introduction}

The effect of the first order of the electron-phonon interaction (EPI) is scattering of free electrons and holes by the optical and acoustic phonons. At the same time there is a second order effect associated with the fact that in some cases phonons can change the energy spectrum of free charge carriers. This phenomenon is called the polaron effect while the charge carrier interacting with phonons is termed polaron. Polaron theory - it is a vast area of solid state physics. Now it is experiencing a new surge of activity. First of all, it is connected with the research of high-temperature superconductors, various multiferroics, manganites with colossal magnetoresistance and quasi-one-dimentional (q1D) systems. In addition, recently appeared the possibility of modeling and simulation of different polaron effects in systems of ultracold atoms in 1D and 2D optical lattices strongly attracted the attention of researchers because it allows by controlled way to change the parameters of interest of the physical system~\cite{1, 2, 3}. The optical lattice with carriers is immersed in a Bose-Einstein condensate resulting in a polaron state formation in form of a carrier dressed in a cloud of coherent phonons - Bogolyubov's excitations~\cite{4, 5}. In this paper, we will only refer to q1D- systems.

Long before the q1D- crystals were obtained experimentally theoretical studies on 1D electron systems showed that the 1D electronic systems differ significantly from the 2D- or 3D- systems. The following three statements could be made to highlight the specific features of 1D- systems~\cite{6}.

a) The 1D metallic system without taking into account the Coulomb interaction of the electrons is unstable with respect to the periodic potential with the wave vector $2{k_F}$. This instability leads to the formation of a self-consistent periodic changes in the electron density and the displacement of lattice atoms and the opening of the gap in the energy spectrum at the Fermi level. In other words, as the temperature decreases lattice distortions should appear with wave number equal to twice the Fermi momentum, and the ground state of a 1D chain of atoms at zero temperature should be dielectric~\cite{7}. Such a self-consistent change in the electron density and the position of the lattice atoms is called a charge density wave (CDW).

b) In the 1D electron system with a half-filled band single-electron excitations are separated from the ground state by a gap at an arbitrarily weak repulsion of electrons~\cite{8}. This assertion is proved for the case when the electron interaction is described within the framework of the Hubbard Hamiltonian. However, there is no reason to believe this statement is unjust for a real Coulomb interaction of electrons. Thus, the Coulomb interaction between the electrons leads to the dielectric (Mott) transition with decreasing temperature.

c) The single-electron states in 1D- system are localized at an arbitrarily weak random potential. Therefore, at low temperatures the conductivity of the system in the lattice with defects can not be metallic~\cite{9}.

All these statements show that, at least for three reasons 1D electron system can be nonmetallic at low temperatures.

In the recent years, the study of the pseudogap in the spectrum of elementary excitations of different q1D- systems has been of great interest. The pseudogap anomalies were observed in a number of experiments such as measurements of optical conductivity, inelastic neutron scattering, angle-resolved photoemission spectroscopy (ARPES)~\cite{10}.

Characteristic features of the ARPES- signal intensity spectra for q1D- compounds with CDW is a shift of the maximum intensity of the spectrum in depth from the Fermi level, and the broadening of the intensity maximum with a stronger energy blur than for the conventional 3D quasiparticles in metals at the Fermi level, for which the maximum is described by a Lorentzian. Out of the inorganic materials ARPES has been applied to blue bronze $K_{0.3}MoO_3$~\cite{11} and to $(TaSe_4)_2I$~\cite{12}.

In the mean field theory the density of states (DOS) for the single-particle excitations in the framework of the Frohlich model is described by inverse square root dependence ${{dN} \mathord{\left/
 {\vphantom {{dN} {dE}}} \right.
 \kern-\nulldelimiterspace} {dE}} = D\left( E \right) \sim {1 \mathord{\left/
 {\vphantom {1 {\sqrt {E - 2\Delta } }}} \right.
 \kern-\nulldelimiterspace} {\sqrt {E - 2\Delta } }}$, but in the spectra of real q1D- compounds with density waves the inverse square root dependence was never observed~\cite{13}. The experimental spectra of the DOS are always blurred in the vicinity of energies $E = 2\Delta $ by the amount more than ${k_B}T$~\cite{14, 15}. One of the reasons is the strong fluctuations of the order parameter and the EPI, which leads to the interaction of the free carriers with the fluctuations and the formation of localized states. The fluctuations, according to modern concepts, also lead to a difference between the temperature of the Peierls transition from a value predicted by mean field theory.

The order parameter, describing the modulation of CDW, is given by $\Delta  = g\left( {2{k_F}} \right)\left\langle {{b_{2{k_F}}} + b_{ - 2{k_F}}^ + } \right\rangle {e^{i2{k_F}x}} = \left| \Delta  \right|{e^{i2{k_F}x}}$, where the brackets $\left\langle {} \right\rangle $ denote thermodynamic average. If a $2\Delta $ gap opens in the electronic spectrum at the Fermi level then the dispersion of the single-particle excitation becomes $E\left( k \right) = {\mathop{\rm sgn}} \varepsilon \left( k \right){\left[ {{\varepsilon ^2}\left( k \right) + {{\left| \Delta  \right|}^2}} \right]^{{1 \mathord{\left/
 {\vphantom {1 2}} \right.
 \kern-\nulldelimiterspace} 2}}}$. The predicted ratio between the value of Peierls gap and the critical temperature is the same as for superconductivity ${{2\Delta } \mathord{\left/
 {\vphantom {{2\Delta } {{k_B}{T_C}}}} \right.
 \kern-\nulldelimiterspace} {{k_B}{T_C}}} = 3.52$. However, according to numerous experimental data ~\cite{13, 16, 17} in the q1D inorganic conductors ${{2\Delta } \mathord{\left/
 {\vphantom {{2\Delta } {{k_B}{T_C}}}} \right.
 \kern-\nulldelimiterspace} {{k_B}{T_C}}} = 8 - 14$ (the ratio depends on the sample compound) and therefore one distinguishes the transition temperature in the mean field theory ${T_{MF}}$ and experimentally determined ${T_P}$. According to modern concepts ${T_P}$ corresponds to the temperature of 3D ordering ${T_{3D}}$, in which the interaction between CDW fluctuations of the order parameter on adjacent 1D chains (i.e., in a direction perpendicular to the direction of maximum conductivity) leads to the appearance of the order parameter correlation in all three dimensions and the formation of 3D CDW. The theoretical justification is the work~\cite{18}, in which Lee, Rice, and Anderson showed that, strictly speaking, the long-range order in the system is absent at any finite temperature, since the correlation function decays exponentially with the distance: $\left\langle {\Delta \left( x \right)\Delta \left( 0 \right)} \right\rangle  \sim \exp \left( {{{ - x} \mathord{\left/
 {\vphantom {{ - x} {\xi \left( T \right)}}} \right.
 \kern-\nulldelimiterspace} {\xi \left( T \right)}}} \right)$. But below ${T_{3D}} \sim {1 \mathord{\left/
 {\vphantom {1 4}} \right.
 \kern-\nulldelimiterspace} 4}{T_{MF}}$ the correlation length $\xi \left( T \right)$ diverges exponentially, so we can assume that at temperatures below ${{{T_{MF}}} \mathord{\left/
 {\vphantom {{{T_{MF}}} 4}} \right.
 \kern-\nulldelimiterspace} 4}$ in the system a Peierls superlattice is formed. Along this strong fluctuations of the order parameter $\Delta $ also exist at temperatures higher than ${T_{3D}}$. The strong fluctuations are correlated until ${T^*} > {T_{3D}}$, and at $T > {T^*}$ the correlation length ${\xi _ \bot }$ becomes shorter than the distance between the chains. The calculation results show that the system exhibits not a gap, but a dip in the DOS. Only at temperatures $T < {{{T_{MF}}} \mathord{\left/
 {\vphantom {{{T_{MF}}} 4}} \right.
 \kern-\nulldelimiterspace} 4}$ the DOS approaches to that obtained in the molecular field approximation. Decrease of the transition temperature, predicted in~\cite{18} - is a result of compromise between two opposite tendencies: on one hand the state with CDW has the lowest energy if $T < {T_{MF}}$, while on the other hand, in the strict 1D- system at the final temperature long-range order is not possible.

Direct experimental manifestations of fluctuations in q1D- conductors are the x-ray reflections blur, corresponding to the superstructure, and the observation of a pseudogap in the optical spectra at temperatures $T > {T_{3D}}$~\cite{13, 17}, as well as the fluctuations of the amplitude mode CDW excitation, directly observed in experiments on femtosecond spectroscopy~\cite{19}.

There is been quite a lot of theoretical work, in which attempts were made to explain the observed anomalies. Three main areas of research can be identified. One of them is based on a picture of the formation of polarons in which shifting and blurring of the maximum of the DOS are explained in the framework of the polarons theory, i.e. the quasiparticles are recognized by mobile polarons with a short coherence length. The interaction with phonons increases the effective mass of the carrier and leads to the appearance of harmonics near the quasiparticle peak at ${E_F}$, instead of a typical Lorentzian, as well as the blurring and displacement of the quasiparticle peak on $\left\langle n \right\rangle \hbar \omega $, where $\left\langle n \right\rangle $ - the average number of phonons interacting with electrons, and $\hbar \omega $ - the characteristic phonon energy. However, in the recent studies for $K_{0.3}MoO_3$~\cite{20}, where the quasiparticle peak with a fine structure was obtained at 80K with a resolution of about 1 meV, have shown that the fine features, as well as a small coherence length of the quasiparticles, recovered from the $k$- dispersion, are better described in the so-called theoretical "ladder" model, in which electron-electron interactions are essential and in the framework of which, the bound states due to the presence of spin and holon excitations are responsible for the peak. Features inherent to these excitations in the ARPES spectra are blurred due to the Gaussian fluctuations and fluctuations of the wave vector on the surface of the crystal. At the same time in $(TaSe_4)_2I$ EPI remains essential and the CDW gap is likely to open in the background of polaron gap that exists at temperatures above ${T_P}$~\cite{21}.

Another direction suggests that the pseudogap phenomena are caused mainly by fluctuations of the short-range order of the CDW type. Quite a long time ago Sadowskii proposed an exactly solvable model of formation of a pseudogap in the 1D- system due to developed fluctuations of short-range order of charge or spin density wave (SDW) type~\cite{22, 23, 24}. This model is of some interest in connection with attempts to explain the pseudogap state of high-Tc cuprates~\cite{25, 26}. In particular, in~\cite{25, 26} a significant generalization of this model was put forward to the case of 2D electron system in a random field of developed spin fluctuations (of antiferromagnetic short-range order).

As a rule, in q1D- systems the impact of thermodynamic fluctuations of the order parameter on the Peierls gap is considered, following Brazovskii~\cite{27, 28}, similar to the influence of static disorder - introducing a random potential with the distribution of the type of Gaussian white noise. Brazovskii first received the blur of optical spectra, bound soliton states and some other features.

In the third approach, phonon effects are neglected and the ifluence of electron correlations, and non-Fermi liquid behavior of q1D- systems are considered in a Tomonaga-Luttinger model~\cite{29, 30}.

\section{\label{sec:pGTB}The polaronic version of the GTB method}
In this paper, we will not use perturbation theory on the EPI. Instead in the framework of the method GTB~\cite{31} (Generalized Tight Binding Approach) we will consider the case of weak and strong EPI. Initially, GTB and it is ab initio version LDA + GTB were proposed to describe the electronic structure of cuprates - high-temperature superconductors, different Mott-Hubbard systems and, in fact, is an implementation of the cluster perturbation theory in terms of the Hubbard  $X$- operators. In~\cite{32} polaron version of GTB method (p-GTB) has been proposed for calculating the electronic structure of strongly correlated systems with strong EPI. This approach can be divided into three main stages: 1) Partition of infinite crystal lattice into a plurality of unit cells (clusters), in our case of 1D chain it will be a clusters with two atoms. An exact diagonalization of intracell Hamiltonian with EPI and finding the energy ${E_p}$ and the many-particle wave functions $\left| p \right\rangle $ of local polarons. 2) Constructing of polaron Hubbard  $X$- operators $X_f^{pq} = \left| p \right\rangle \left\langle q \right|$ on the basis of local multi-electron and multi-phonon eigenstates found in the first step. Indices "$p$" and "$q$" contain a set of quantum numbers characterizing the state of the system. Calculation of matrix elements of the creation and annihilation operators for both electrons and phonons in this basis will allow to write down on-site single-electron and phonon operators as a linear combination of the Hubbard operators; quasifermion for electrons, and quasiboson for phonons. 3) In general, the multi-band model with electron-electron and EPI interactions is written as a generalized Hubbard model in the representation  $X$- operators with a set of local polaron states and intercell hopping. The $X$- operators representation allows us to take into account the strong correlation and electron-phonon coupling in the zeroth order approximation. The dispersion of the band structure of Fermi and Bose excitations occur due to the itercell hopping. An important new aspect of the theory developed is the dependence of the quasiparticle dispersion law on the occupation numbers of the local states. In our case, the temperature will change occupation numbers of different multi-electron terms and multi-phonon levels, which may lead to a strong dependence of the polaron dispersion on temperature.

\section{\label{sec:model}The minimal model}
At the Peierls transition the EPI affects not only electronic, but also the phonon system (Kohn anomaly). Therefore, for the consistent treatment of the Peierls transition, it is necessary to introduce the Hamiltonian, describing electrons, phonons and their interaction.

In the tight binding model, we can assume that the displacement of the ions from their equilibrium positions change only the resonance integrals $t$ (hopping integrals). Then, the electron Hamiltonian taking into account the motion of the ions is given by
\begin{eqnarray}
\begin{array}{l}
H = \sum\limits_i {\frac{{p_i^2}}{{2M}}}  + \frac{1}{4}M\omega _0^2\sum\limits_i {{{\left( {{u_{i + 1}} - {u_i}} \right)}^2}} \\
 - t\sum\limits_{i,\sigma } {\left( {a_{i + 1,\sigma }^ + {a_{i,\sigma }} + a_{i,\sigma }^ + {a_{i + 1,\sigma }}} \right)} \\
 + t'\sum\limits_{i,\sigma } {\left( {{u_{i + 1}} - {u_i}} \right)\left( {a_{i + 1,\sigma }^ + {a_{i,\sigma }} + a_{i,\sigma }^ + {a_{i + 1,\sigma }}} \right)},
\end{array}
\label{eq_H1}
\end{eqnarray}
where ${p_i}$ and ${u_i}$ - momentum operators and the displacement of an atom at site $i$, $t'$ - a derivative of the resonance integral with respect to the interatomic distance, and $M$ - the mass of the ions. Further, one usually transforms to the phonon representation and obtains a Frohlich Hamiltonian, but we prefer to work in the presentation of "naked" Einstein phonons. Let us introduce the creation and annihilation operators of the quantum excitation $i$-th oscillator, i.e. $b_i^ + $ and ${b_i}$:
${u_i} = \sqrt {\frac{1}{{2M{\omega _0}}}} \left( {{b_i} + b_i^ + } \right)$, ${p_i} = \frac{1}{i}\sqrt {\frac{{M{\omega _0}}}{2}} \left( {{b_i} - b_i^ + } \right)$. \\
As a result we obtain
\begin{eqnarray}
\begin{array}{l}
H = {\omega _0}\sum\limits_i {\left( {b_i^ + {b_i} + \frac{1}{2}} \right)}  - \frac{{{\omega _0}}}{4}\sum\limits_i {\left( {{b_i} + b_i^ + } \right)\left( {{b_{i + 1}} + b_{i + 1}^ + } \right)} \\
 - t\sum\limits_{ij,\sigma } {a_{i,\sigma }^ + {a_{j,\sigma }}}  + \sum\limits_{ijk,\sigma } {\left( {\lambda _{jk}^ib_i^ +  + \lambda _{kj}^{i * }{b_i}} \right)a_{j,\sigma }^ + {a_{k,\sigma }}},
\end{array}
\label{eq_H2}
\end{eqnarray}
where $\lambda $ - is a parameter of EPI. Below we will use, as is customary, the dimensionless parameter of EPI and distinguish two cases of weak ($\lambda  = 0.05$) and strong ($\lambda  = 0.4$) EPI.

Single-electron annihilation and creation operators in a cell $f$ with a spin projection $\sigma$ can be expressed as a linear combination of $X$- operators
${a_{f\sigma }} = \sum\limits_{pq} {\left| p \right\rangle \left\langle {p\left| {{a_{f\sigma }}} \right|q} \right\rangle \left\langle q \right|}  = \sum\limits_{pq} {{\gamma _\sigma }\left( {pq} \right)X_f^{pq}} $,
$a_{f\sigma }^ +  = \sum\limits_m {\gamma _\sigma ^ * \left( {pq} \right)X_f^{ + pq}} $.
Alternatively, as the number of different root vectors $\left( {pq} \right)$ is finite, the vectors can be numbered and each vector could be indexed with number $m$, so
${a_{f\sigma }} = \sum\limits_m {{\gamma _\sigma }\left( m \right)X_f^m} $, $a_{f\sigma }^ +  = \sum\limits_m {\gamma _\sigma ^ * \left( m \right)X_f^{ + m}} $.

In the $X$- operators representation the Hamiltonian (2) has the form
\begin{eqnarray}
\begin{array}{l}
H = \sum\limits_{f,p} {\left( {{E_p} - {n_p}\mu } \right)X_f^{pp}}
+ \sum\limits_{f \ne g} {\sum\limits_{m,n} {\left( {t_{fg}^{mn} + M_{fg}^{mn}} \right){X^ + }_f^mX_g^n} },
\end{array}
\label{eq_H3}
\end{eqnarray}
where $\mu $ - the chemical potential, $t_{fg}^{mn} = \sum\limits_\sigma  {{t_{fg}}\gamma _\sigma ^*\left( m \right){\gamma _\sigma }\left( n \right)} $ contains the integrals of intercluster hopping ${t_{fg}}$, and $\Lambda _{fg}^{mn} = \sum\limits_\sigma  {{\Lambda _{fg}}\gamma _\sigma ^*\left( m \right){\gamma _\sigma }\left( n \right)} $ contains the terms of intercell EPI ${\Lambda _{fg}}$.

To obtain the dispersion relations of quasiparticle excitations we use the equations of motion for the polaron matrix Green's function: ${D_{mn}}\left( {k,\omega } \right) = \left\langle {\left\langle {{X_k^m}}
 \mathrel{\left | {\vphantom {{X_k^m} {X_k^{ + n}}}}
 \right. \kern-\nulldelimiterspace}
 {{X_k^{ + n}}} \right\rangle } \right\rangle $, related with one-electron Green's function ${G_\sigma }\left( {k,\omega } \right) = \left\langle {\left\langle {{{a_{k\sigma }}}}
 \mathrel{\left | {\vphantom {{{a_{k\sigma }}} {a_{k\sigma }^ + }}}
 \right. \kern-\nulldelimiterspace}
 {{a_{k\sigma }^ + }} \right\rangle } \right\rangle $ by the ratio
${G_\sigma }\left( {k,\omega } \right) = \sum\limits_{m,n} {{\gamma _\sigma }\left( m \right)\gamma _\sigma ^ * \left( n \right){D_{mn}}\left( {k,\omega } \right)} $.

The spectral density of single-particle excitations is expressed through Fermi-particle Green's function
${A_\sigma }\left( {k,\omega } \right) =  - \frac{1}{\pi }\sum\limits_{mn} {{\gamma _\sigma }\left( m \right)\gamma _\sigma ^ * \left( n \right){\mathop{\rm Im}\nolimits} {D_{mn}}\left( {k,\omega  + i\delta } \right)}  =  - \frac{1}{\pi }{\mathop{\rm Im}\nolimits} G\left( {k,\omega  + i\delta } \right)$
and the density of the single-particle states for a given spin projection (${N_k}$ - normalization factor) is given by
${N_\sigma }\left( \omega  \right) = \frac{1}{{{N_k}}}\sum\limits_k {{A_\sigma }\left( {k,\omega } \right)} $.

Dyson generalized equation can be written for the $\hat D$ Green's function~\cite{32, 33}
\begin{eqnarray}
{\hat D_k}\left( \omega  \right) = \left[ {\hat G_0^{ - 1}\left( \omega  \right) - {{\hat P}_k}\left( \omega  \right)\left( {{{\hat t}_k} + {{\hat \Lambda }_k}} \right) + {{\hat \Sigma }_k}\left( \omega  \right)} \right]{\hat P_k}\left( \omega  \right).
\label{eq_Dyson}
\end{eqnarray}
Here ${\hat \Sigma _k}\left( \omega  \right)$ and ${\hat P_k}\left( \omega  \right)$ - respectively the mass and power operators, ${\hat G_0}\left( \omega  \right)$ - local intracell propagator, $\hat t_k^{mn} = \sum\limits_\sigma  {\gamma _\sigma ^ * \left( m \right){\gamma _\sigma }\left( n \right){t_k}} $, where ${t_k}$ - the Fourier image of the intercluster hopping and similarly $\hat \Lambda _k^{mn} = \sum\limits_\sigma  {\gamma _\sigma ^ * \left( m \right){\gamma _\sigma }\left( n \right){\Lambda _k}} $.

In the Hubbard-1 approximation, the exact structure of Green's function (4) is preserved, but the mass operator is assumed to be zero, and the force operator $\hat P_k^{mn}\left( \omega  \right) = {\delta _{mn}}F\left( m \right)$, where $F\left( m \right) \equiv F\left( {pq} \right) = \left\langle {{X^{pp}}} \right\rangle  + \left\langle {{X^{qq}}} \right\rangle $ - is the filling factor referred to as the end factor in the diagram technique for $X$- operators~\cite{34}. From equation (4) we obtain $\hat D_k^{ - 1} = \hat D_0^{ - 1} + {\hat t_k} + {\hat \Lambda _k}$. Here $\hat D_0^{mn} = {{{\delta _{mn}}F\left( m \right)} \mathord{\left/
 {\vphantom {{{\delta _{mn}}F\left( m \right)} {\left( {\omega  - \Omega \left( m \right)} \right)}}} \right.
 \kern-\nulldelimiterspace} {\left( {\omega  - \Omega \left( m \right)} \right)}}$, where $\Omega \left( m \right) \equiv \Omega \left( {pq} \right) = {E_p} - {E_q}$. In~\cite{32} it is shown that the term ${\hat \Lambda _k}$ is much smaller than ${\hat t_k}$ (${\Lambda  \mathord{\left/
 {\vphantom {\Lambda  t}} \right.
 \kern-\nulldelimiterspace} t} \sim 0.01$) and therefore does not contribute significantly to the dispersion relations for the quasiparticle excitations.

\section{\label{sec:Band_structure}Polaronic band structure in pGTB method}
Cluster perturbation theory has been successfully used for the Hubbard model~\cite{35, 36} and is exact in the limit of the electron-electron interaction $U = 0$ and $U \to \infty $. In our case, the electron correlations are not considered and just the first case is realized, so the problems associated with the artificial doubling of crystal lattice period, does not occur. This is clearly seen from Fig.~1, which shows the results of our cluster calculations in the absence of the EPI. At $\lambda  = 0$ Hamiltonian (1) is exactly diagonalized in the $k$- space. The electronic band structure is of a metallic type and, as it can be seen in the DOS (Fig.~1(b)) does not exhibit any dip related to the artificial doubling of the period of the 1D chain of atoms, and the spectral density $A\left( {k,\omega } \right)$ at the point $k = \pi/2$ at the energy of chemical potential is given by a Lorentz peak (Fig.~1(c)). The band structure is independent of temperature.

If the system possesses a long-range order (for example, the SDW- in antiferromagnets or CDW- type) in the spectrum of elementary excitations (dielectric) gap opens. Thus, in our calculation for $T = 0$ and $\lambda  \ne 0$ the energy gap ${E_g} = 2\Delta $ opens in spectrum at the Fermi level (Fig.~2(a) and 3(a)). The system becomes dielectric. The width of the gap is determined by the size of the EPI. In either case of weak and strong EPI in the spectral density there are two peaks corresponding to "Bogolyubov" quasiparticles (Fig.~2(a''') and 3(a''')). Thus, regardless to the value of EPI the ground state of the system has a qualitatively the same form - a true dielectric gap caused by long-range order of the CDW-type.

However, with increasing temperature, these two cases are fundamentally different behavior. In the case of a weak EPI the overlap of the spectral weight of the two peaks occurs at $T = T'$ (Fig.~ 2(b''')), and when $T > T'$ they merge into one Lorentzian peak (Fig.~2(c''')), typical for a normal metal (Fermi fluid), and the dielectric gap in the DOS vanishes (Fig.~2(b'') and 2(c'')).

A qualitatively different change of the band structure with increasing temperature can be observed in the case of strong EPI. Fig.~3(b) and 3(c) demonstrated basic polaron effect - bands splitting into polaron subbands and formation of polarons themselves - Fermi type excitations in the system. This is a manifestation of the hybridized state of quasiparticles of the Fermi type and local phonon Franck-Condon resonances~\cite{37, 38}. It is important to note that polaron effects occur with increasing temperature. Because of the strong EPI the substantial renormalization of the single-particle excitations occurs. A significant part of the spectral weight is redistributed between the peak of the coherent quasiparticle excitations and the incoherent part of the spectrum located below on the energy scale. The incoherent part occurred due to the advent of the vibronic satellites. Calculations were performed for the number of phonons ${N_{ph}} = 5$ in the cluster, and therefore in Fig.~3(b''') and 3(c''') one can see five Franck-Condon vibronic satellites resulting from Lorentz broadening of the resonance lines with the width of the Lorentzian $\delta $. At a certain temperature $T = T''$ there is an overlap of the spectral weight of the two peaks (Fig.~3(b''')), but due to the strong reduction of the spectral weight of the coherent quasiparticle peak so-called hidden Fermi surface is opening. Since the ARPES experiments directly measured product $f\left( \omega  \right)A\left( {k,\omega } \right)$, where $f\left( \omega  \right) = {\left[ {\exp ({\omega  \mathord{\left/
 {\vphantom {\omega  T}} \right.
 \kern-\nulldelimiterspace} T}) + 1} \right]^{ - 1}}$ - the Fermi distribution function, the characteristic features of the ARPES-signal spectra intensity for q1D- compounds with CDW is a shift of the spectrum intensity maximum below the Fermi level, and the broadening with a stronger (Gaussian) energy blurring, in comparison against conventional 3D quasiparticles in metals at the Fermi level, for which the maximum is described by a Lorentzian. Despite the fact that the system transforms to the metallic state the DOS still exhibits a dip even at high temperatures $T \gg T''$ (Fig.~3(c'')). All calculations were performed deliberately in the Hubbard-1 approximation to eliminate the attenuation of quasiparticle excitations and the gap blurring with increasing temperature.

In Fig.~4, the solid line shows the temperature dependence of ${{{E_g}} \mathord{\left/
 {\vphantom {{{E_g}} {2{\Delta _{MF}}}}} \right.
 \kern-\nulldelimiterspace} {2{\Delta _{MF}}}}$, where $2{\Delta _{MF}}$ - the gap value in the mean-field theory, as a function of the normalized temperature ${T \mathord{\left/
 {\vphantom {T {{T_{MF}}}}} \right.
 \kern-\nulldelimiterspace} {{T_{MF}}}}$ for two cases of weak (a) and strong (b) EPI. It is seen that $T'$ and $T''$ less than ${T_{MF}}$. The dashed line in Fig.~4(b) shows the behavior of the effective gap ${{E_g^{eff}} \mathord{\left/
 {\vphantom {{E_g^{eff}} {2{\Delta _{MF}}}}} \right.
 \kern-\nulldelimiterspace} {2{\Delta _{MF}}}}$, defined on the level of the DOS which is $e$ times greater than the DOS at the Fermi level $D\left( {{{E_g^{eff}} \mathord{\left/
 {\vphantom {{E_g^{eff}} 2}} \right.
 \kern-\nulldelimiterspace} 2}} \right) = e \cdot D(0)$. It is clearly seen that gap $E_g^{eff}$ exists in the spectrum at arbitrarily high temperatures and reaches a constant value $ \approx {{{E_g}\left( 0 \right)} \mathord{\left/
 {\vphantom {{{E_g}\left( 0 \right)} 2}} \right.
 \kern-\nulldelimiterspace} 2}$ with increasing temperature. This behavior is consistent with the results of~\cite{39}, where a certain effective gap ${\Delta _{eff}}$ in the $K_{0.3}MoO_3$ spectrum was determined experimentally, that exists at $T > {T_P}$.

\section{\label{sec:Concl}Conclusions}
Thus, on the basis of our calculations and comparison between the two extreme cases of the EPI it can be concluded that in the q1D- systems with strong EPI in addition to CDW gap, responsible for the true dielectric ground state, there is a polaron gap or a gap of polaron origin responsible for their pseudogap behavior.

Besides q1D- compounds, the pseudogap effects are observed in manganites with colossal magnetoresistance and high-temperature superconducting cuprates. Understanding the nature of this phenomenon within the framework of the polaron approach remains a topic of constant discussions~\cite{40, 41, 42, 43, 44, 45, 46, 47}.

\begin{acknowledgments}
The authors are thankful to D. Maksimov for assistance in article preparation.

This work was supported by the Russian Foundation for Basic Research, project nos. 17-02-00826, 17-02-00052, 16-02-00507, 16-02-00098, by the Council of the President of the Russian Federation for Support of Young Scientists and Leading Scientific Schools, project no. SP-1844.2016.1, NSh-7559.2016.2, and RFBR and Krasnoyarsk Regional Science Foundation joint projects nos. 16-42-243048, 16-42-240746, 16-43-240505
\end{acknowledgments}

\nocite{*}


\newpage

\begin{figure*}
\begin{center}
$\begin{array}{cc}
 \includegraphics[angle=0,width=0.40\columnwidth]{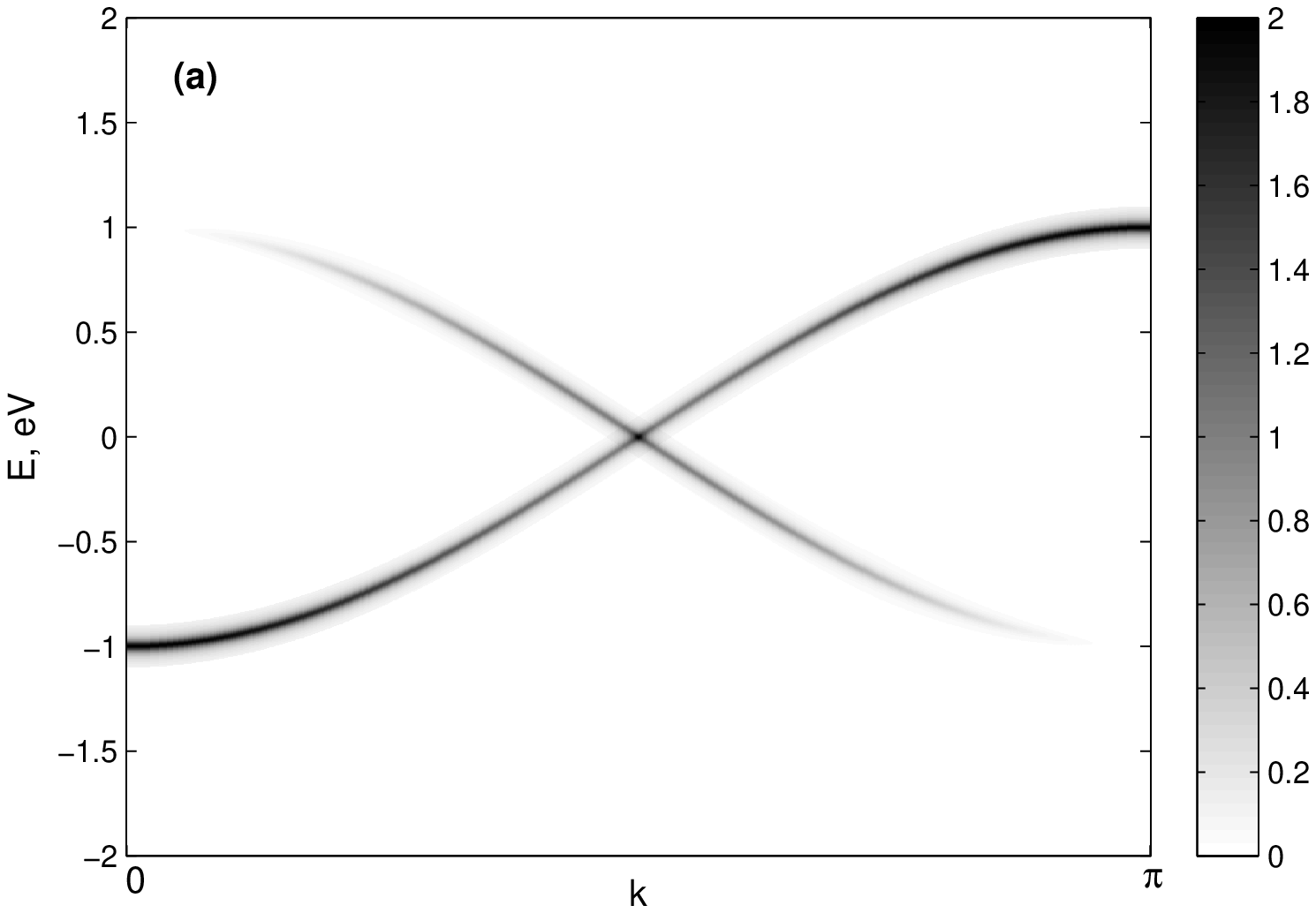}
 \includegraphics[angle=0,width=0.33\columnwidth]{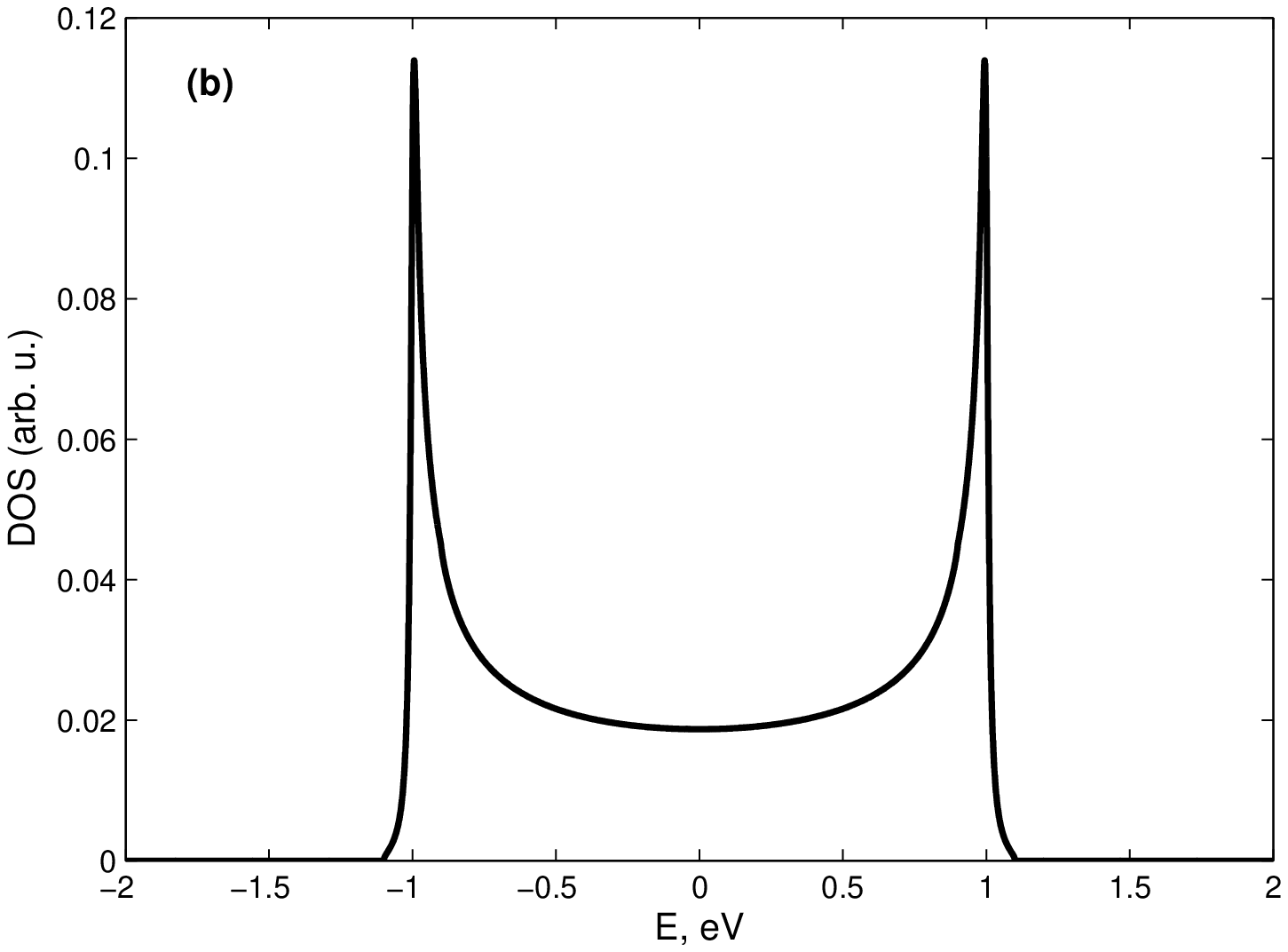}
 \includegraphics[angle=0,width=0.33\columnwidth]{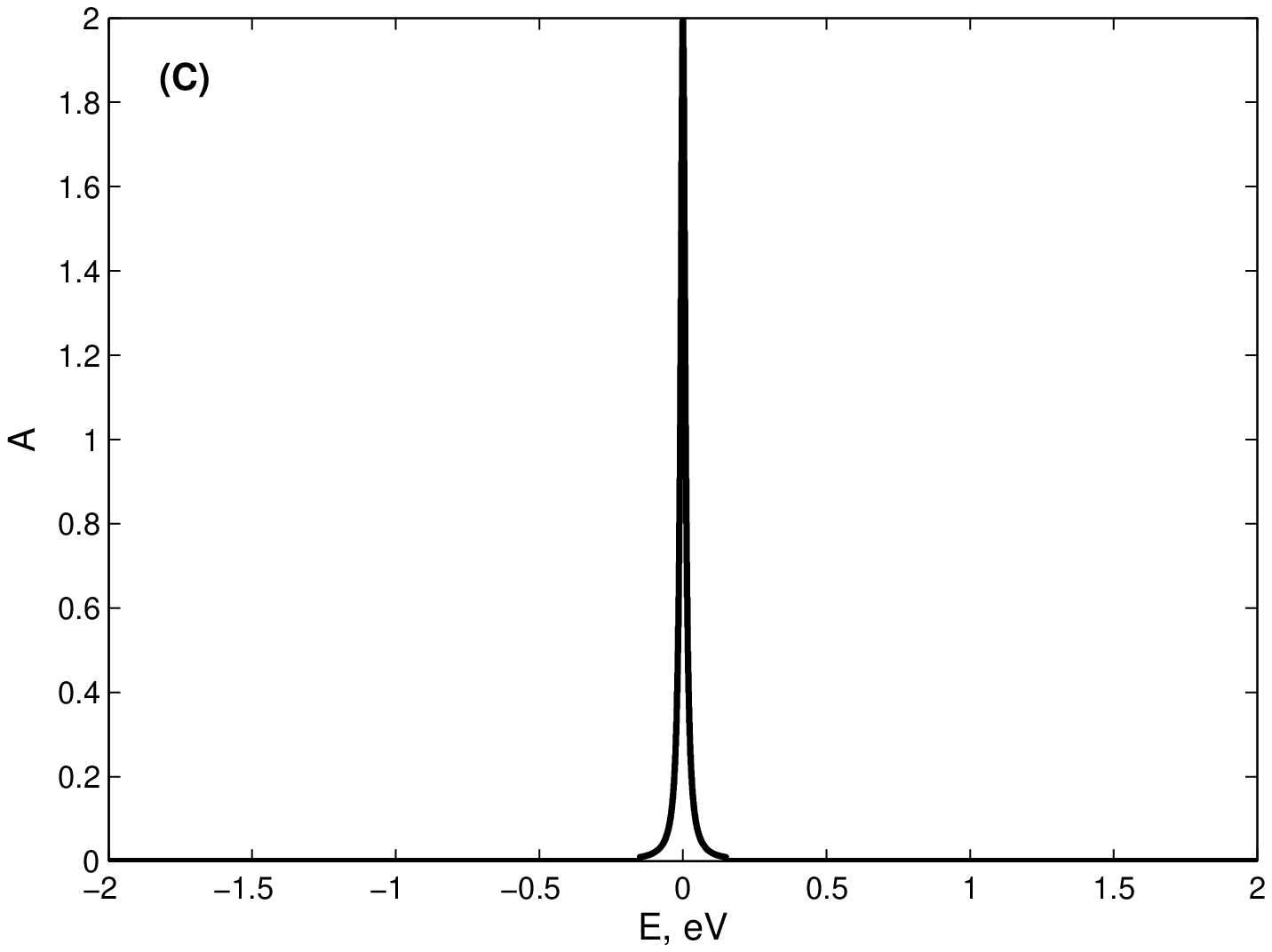}
\end{array}$
\caption{\label{fig1} The electronic structure (band structure, density of states and the spectral density at the point with the wave vector $k = \pi/2$ at the Fermi level) in the absence of EPI. All calculations were performed for the following values of parameters: $\lambda  = 0$, $t = 1$eV, $\delta  = 0.03$eV.}
\end{center}
\end{figure*}
{\sloppy

}
\newpage

\begin{figure*}
\begin{center}
$\begin{array}{cc}
 \includegraphics[angle=0,width=0.35\columnwidth]{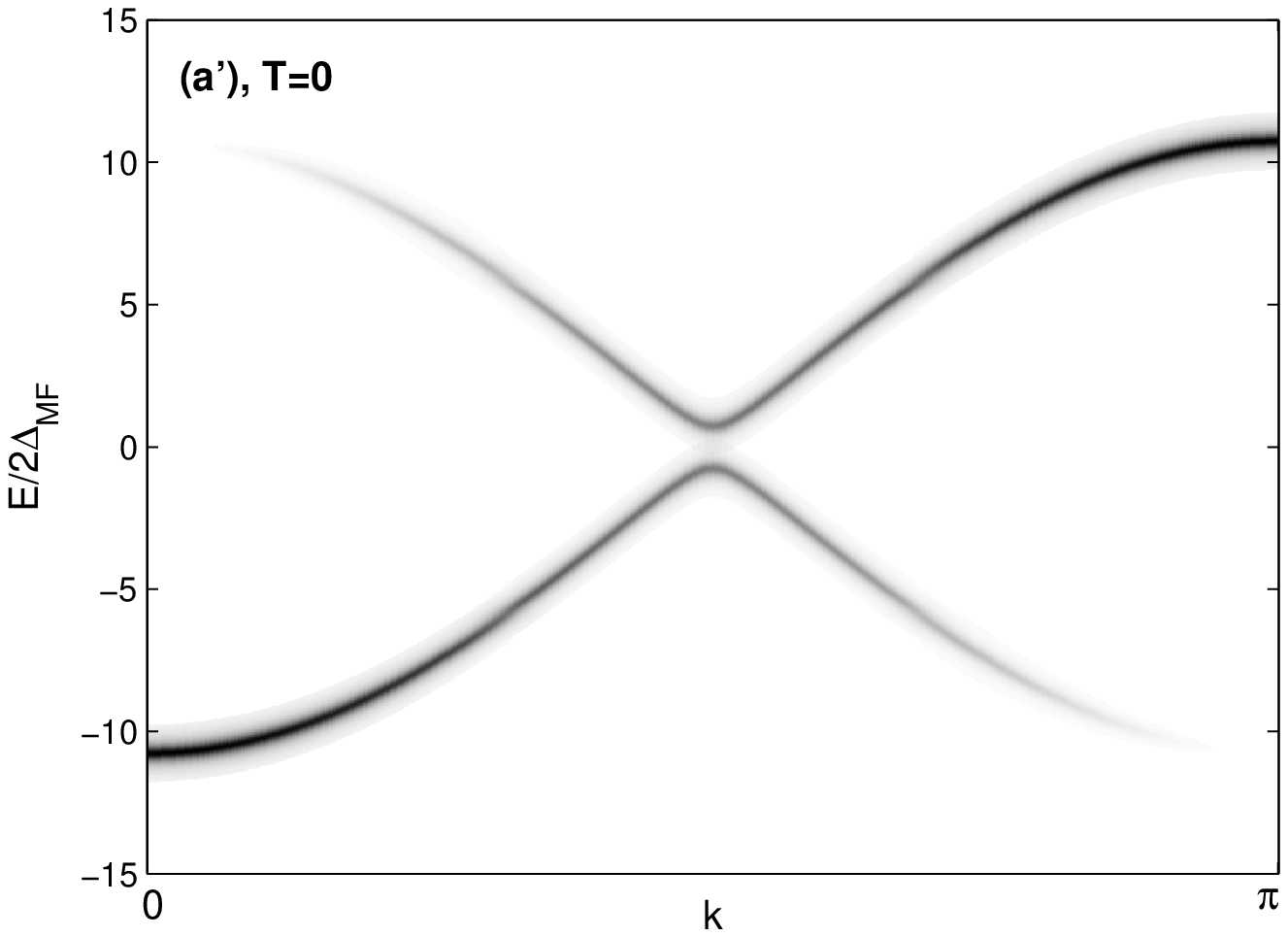}
 \includegraphics[angle=0,width=0.35\columnwidth]{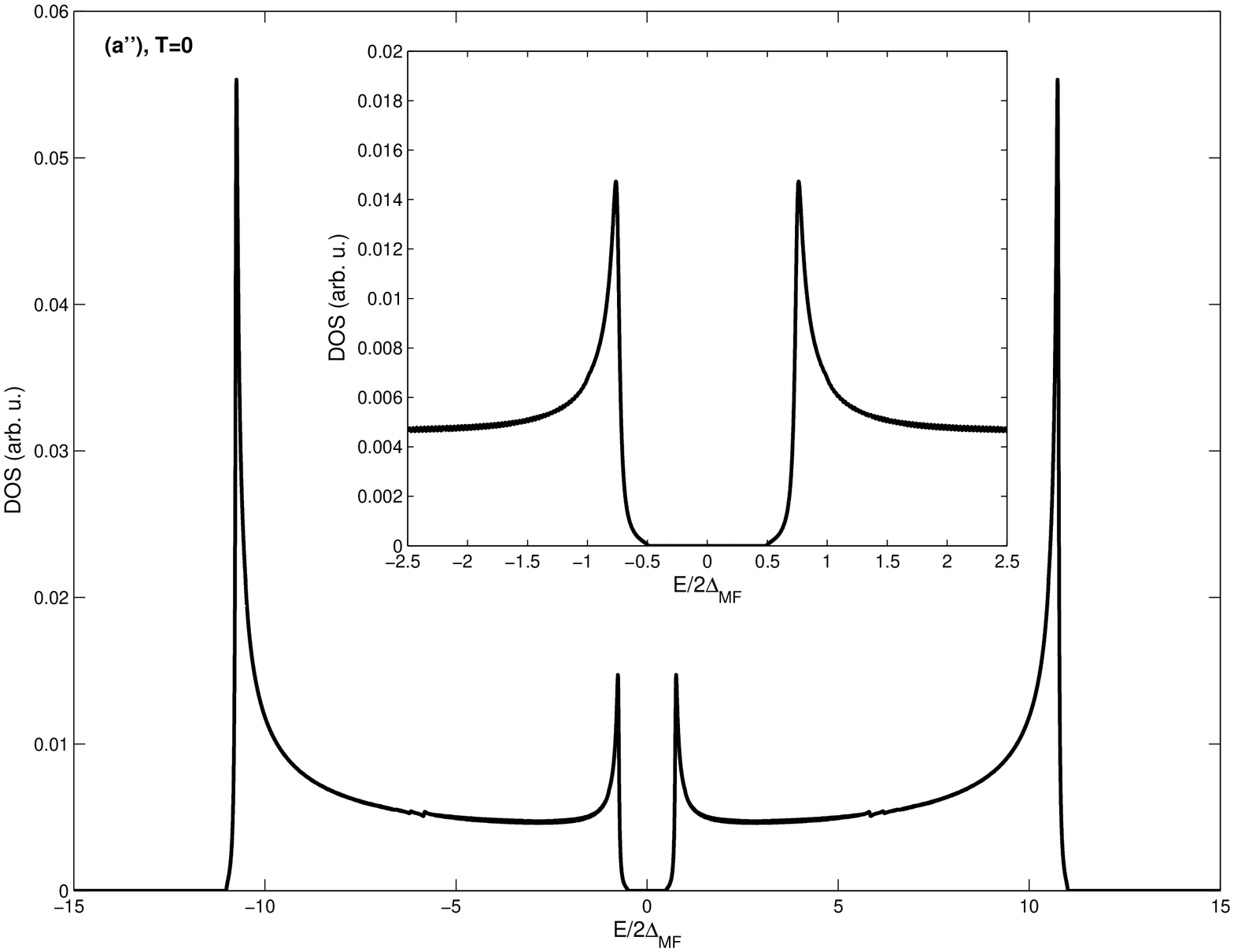}
 \includegraphics[angle=0,width=0.35\columnwidth]{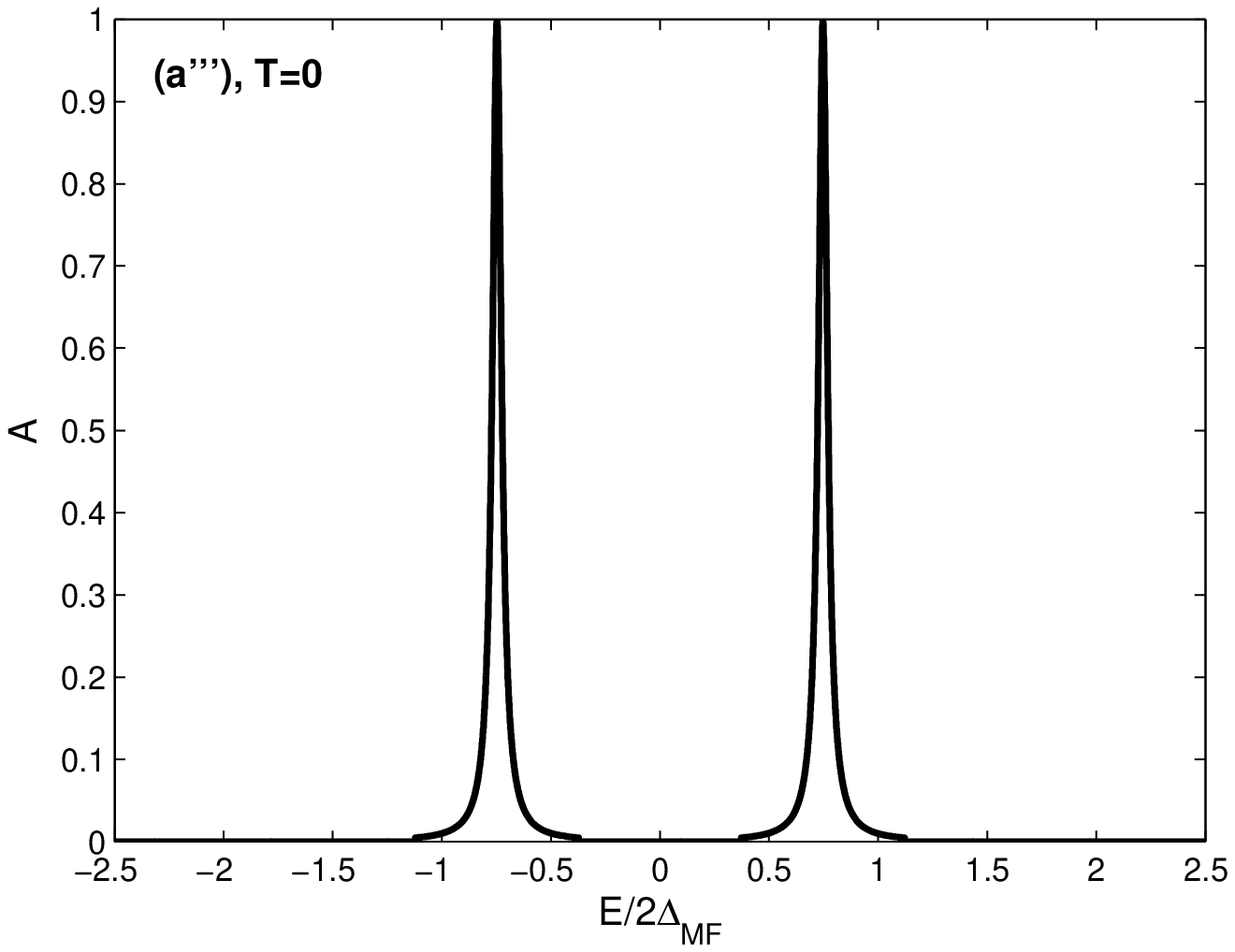}\\
 \includegraphics[angle=0,width=0.35\columnwidth]{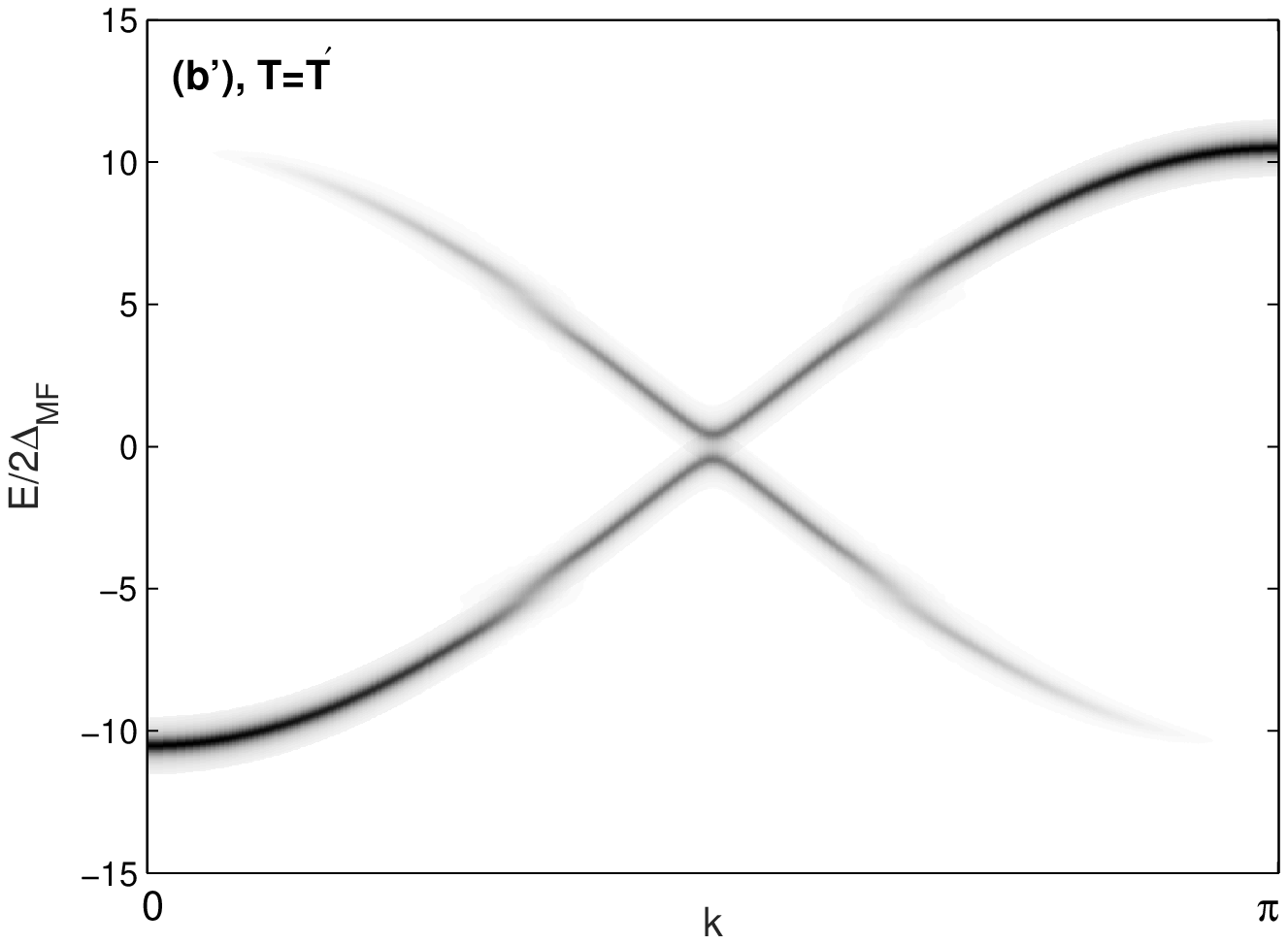}
 \includegraphics[angle=0,width=0.35\columnwidth,height=0.26\columnwidth]{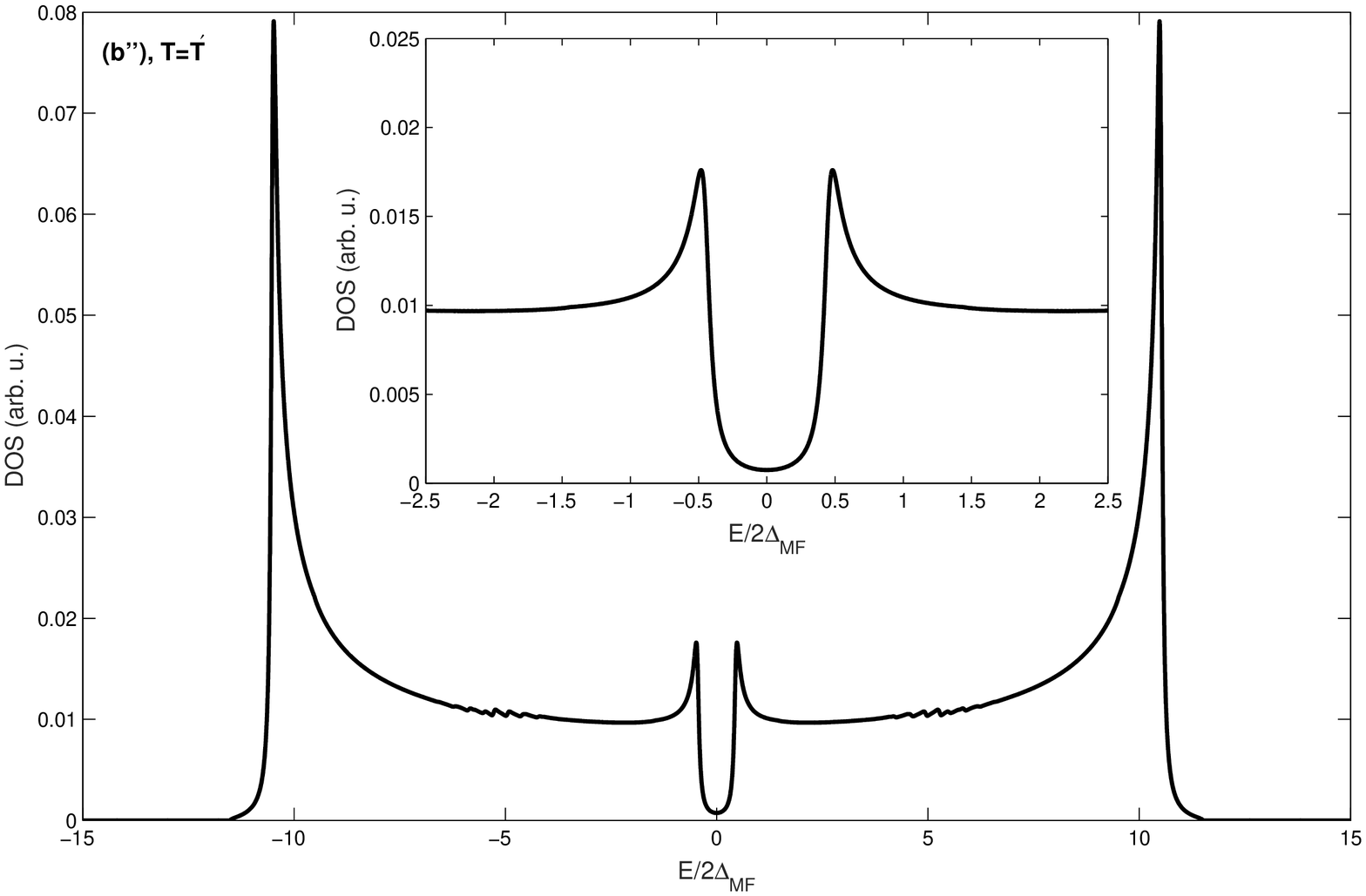}
 \includegraphics[angle=0,width=0.35\columnwidth]{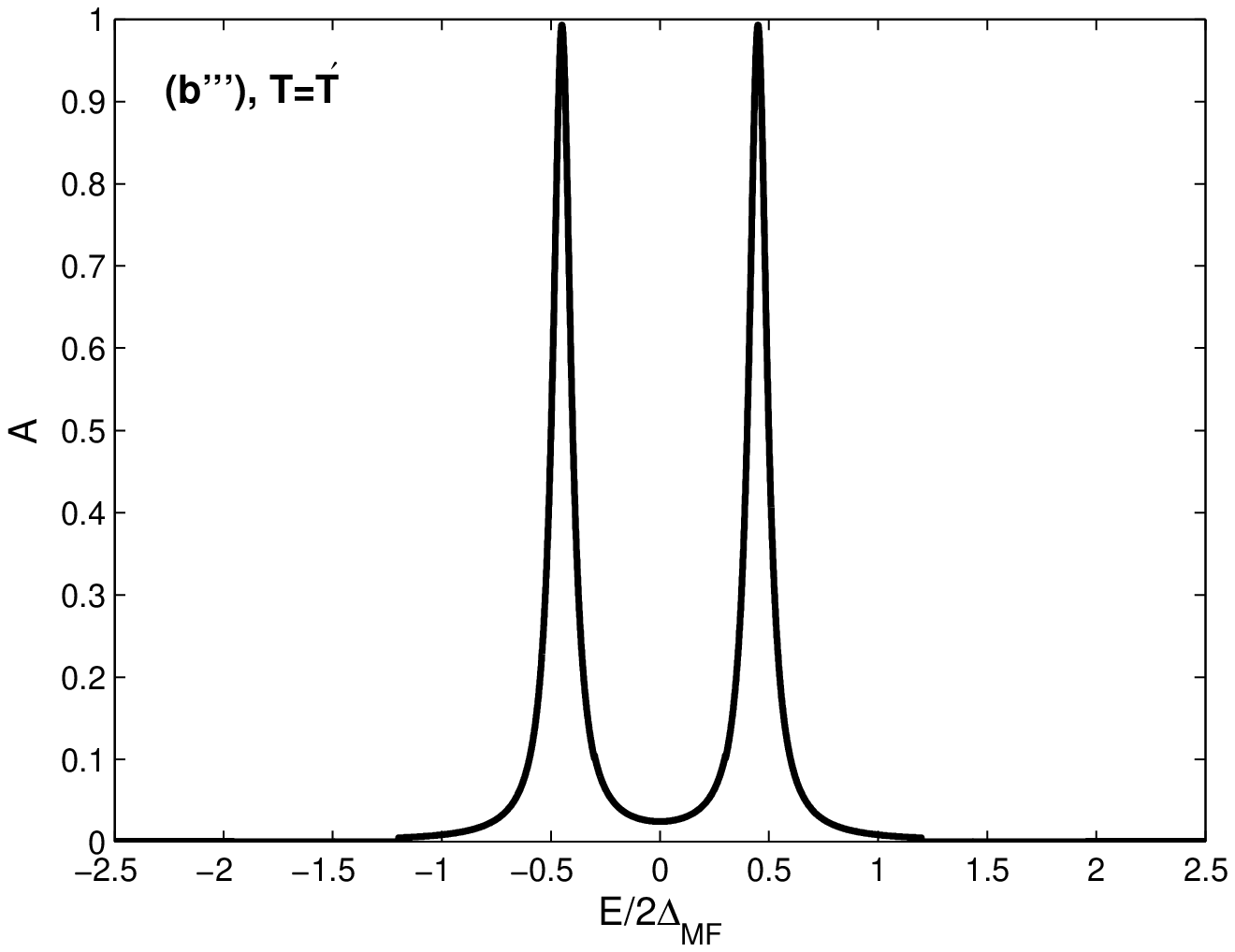}\\
 \includegraphics[angle=0,width=0.35\columnwidth]{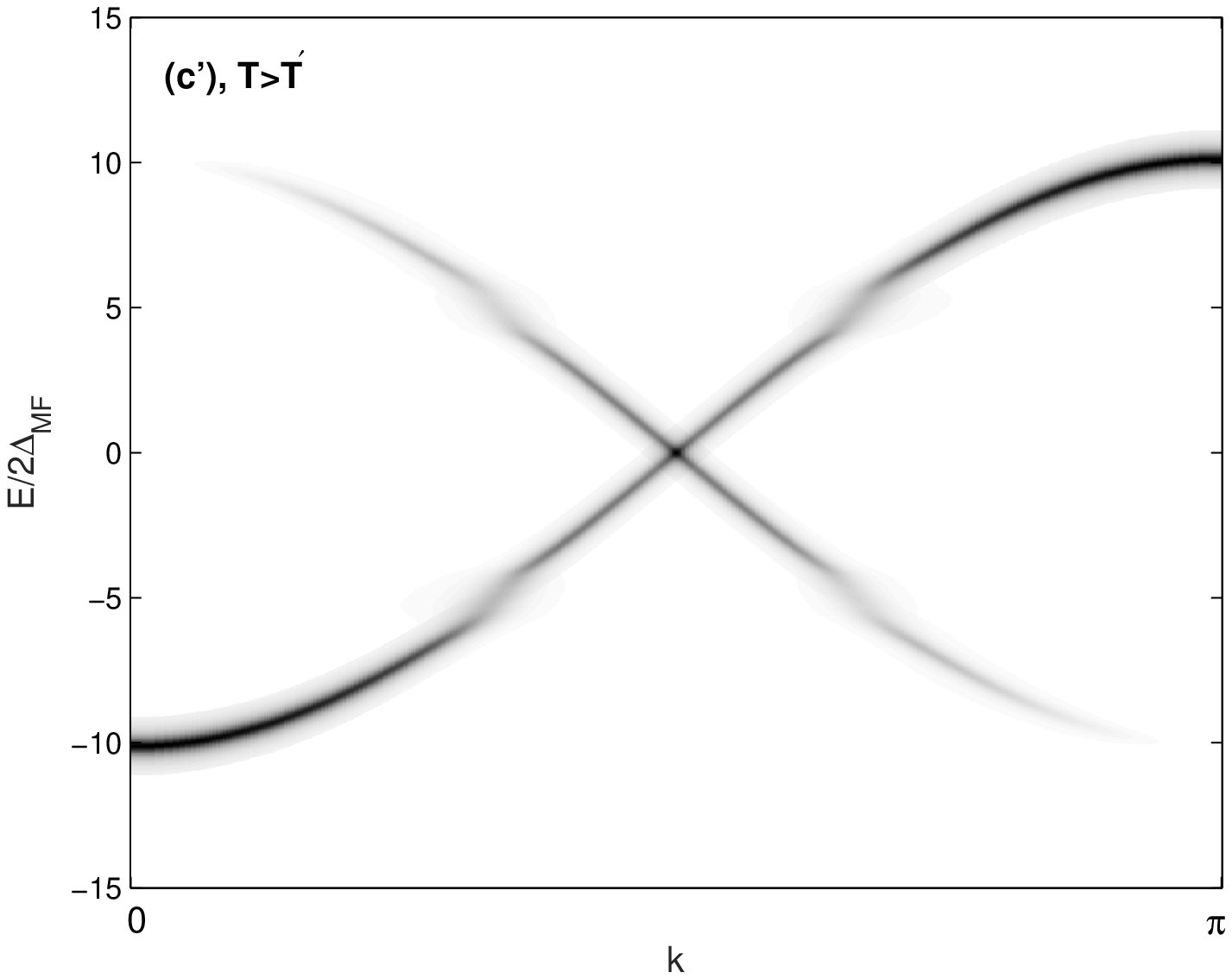}
 \includegraphics[angle=0,width=0.35\columnwidth]{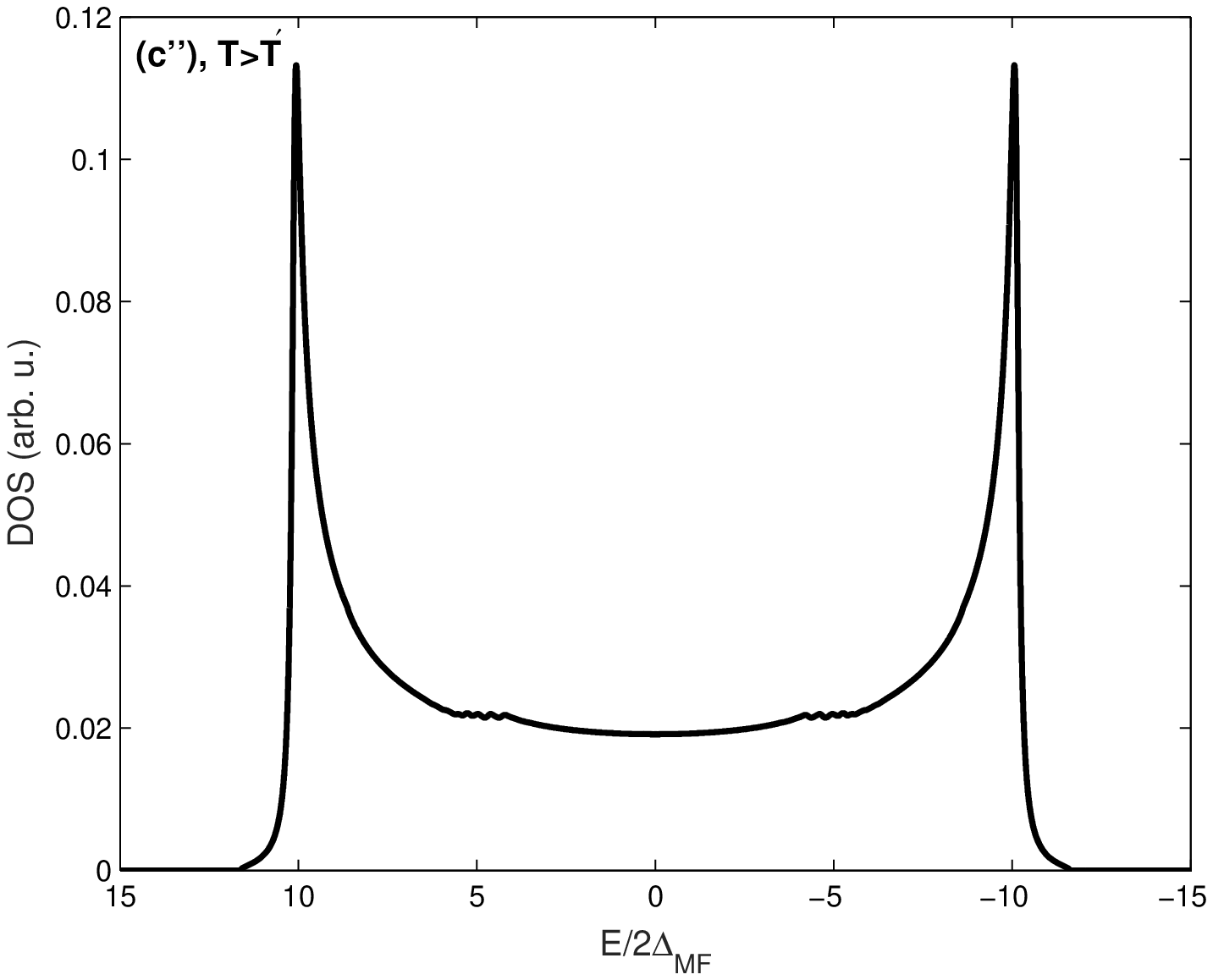}
 \includegraphics[angle=0,width=0.35\columnwidth]{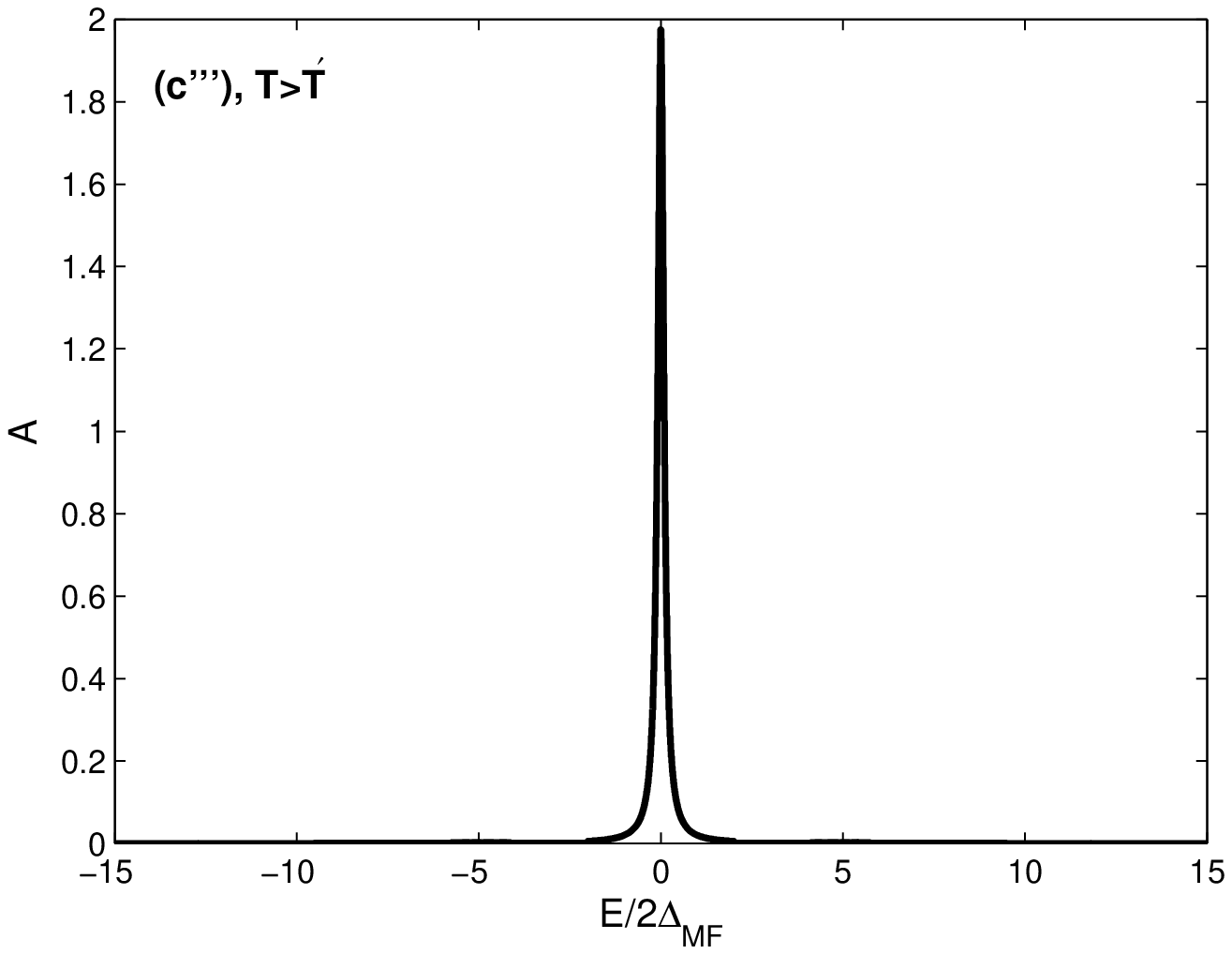}
 \end{array}$
\caption{\label{fig2} Electronic structure (band structure, density of states and the spectral density at the point with the wave vector $k = \pi/2$ at the chemical potential level) in the case of a weak EPI calculated for the three values of temperature (a) $T = 0$, (b) $T = T'$ and (c) $T > T'$. All calculations were performed for the following values of parameters: ${\omega _0} = 0.05$eV, $\lambda  = 0.05$, $t = 1$eV, $\delta  = 0.02 - 0.05$eV.}
\end{center}
\end{figure*}
{\sloppy

}
\newpage

\begin{figure*}
\begin{center}
$\begin{array}{cc}
 \includegraphics[angle=0,width=0.35\columnwidth]{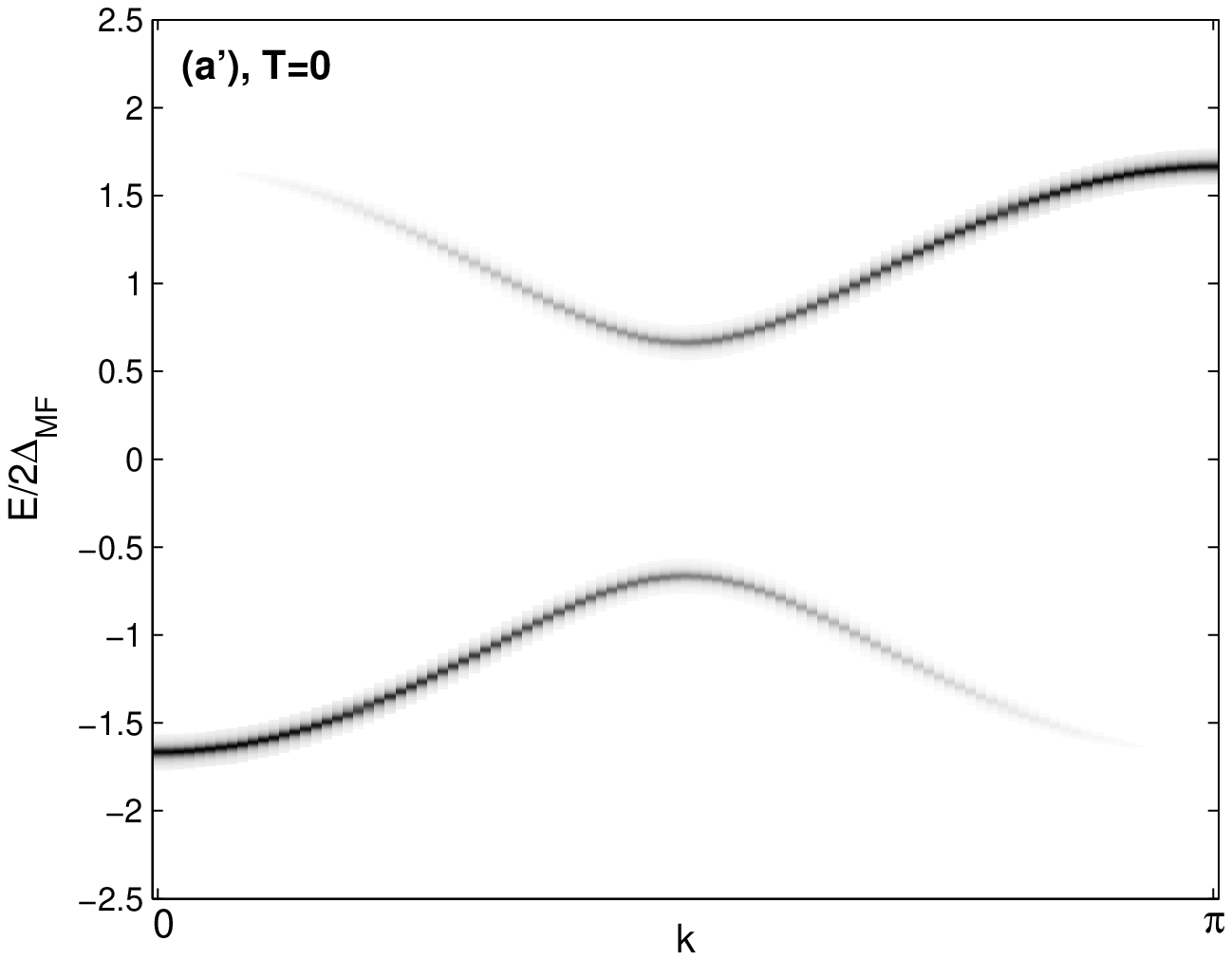}
 \includegraphics[angle=0,width=0.35\columnwidth,height=0.27\columnwidth]{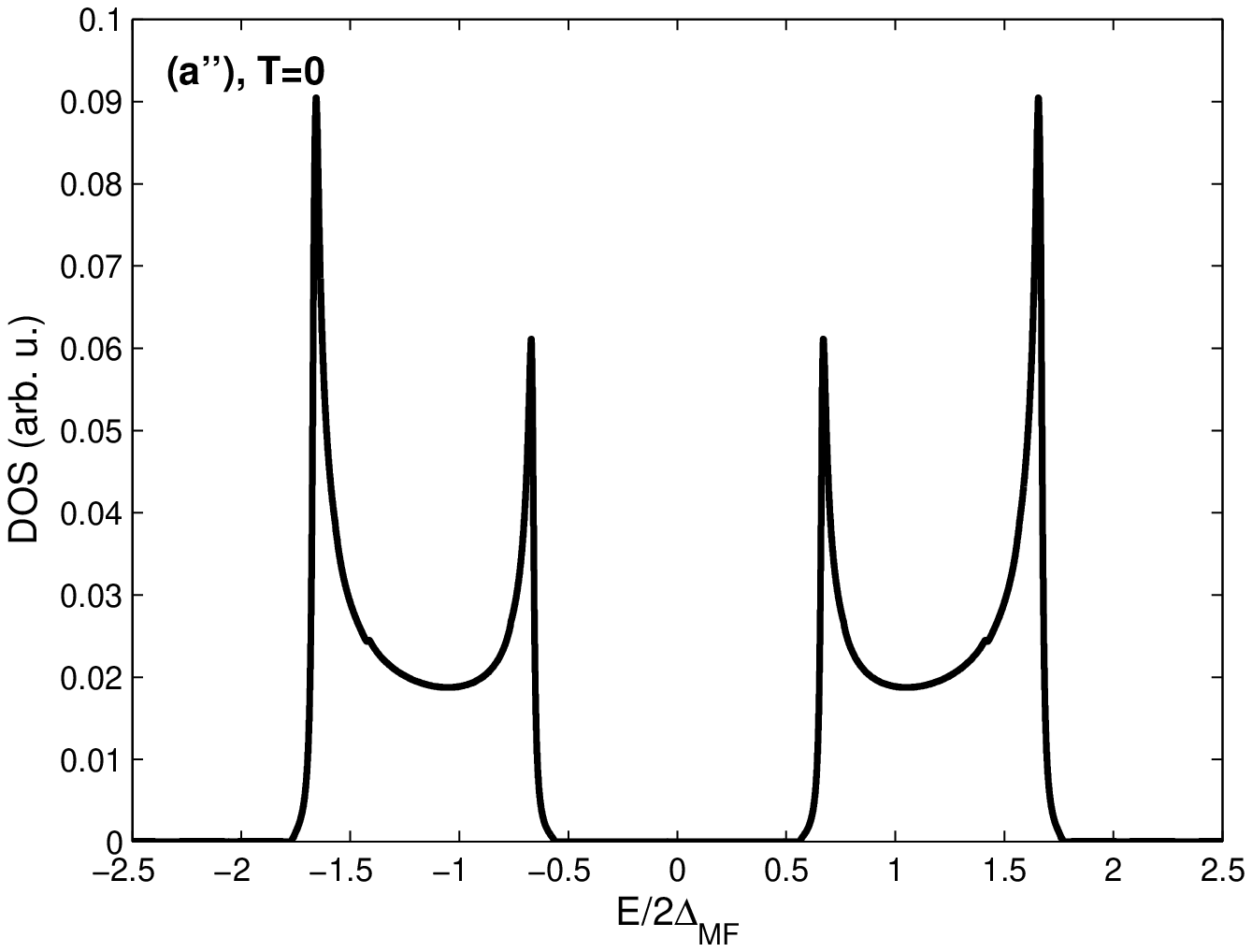}
 \includegraphics[angle=0,width=0.35\columnwidth]{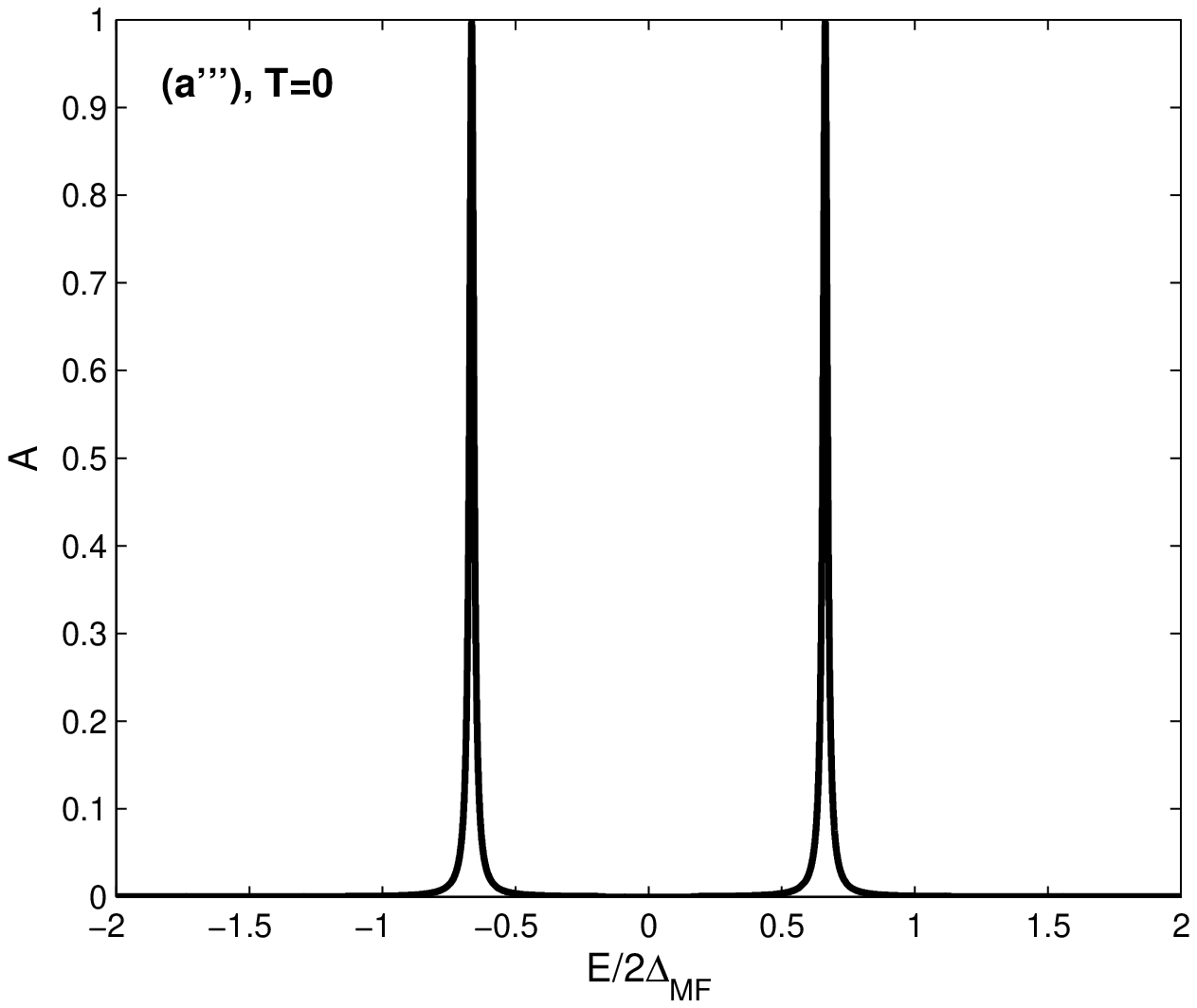}\\
 \includegraphics[angle=0,width=0.35\columnwidth]{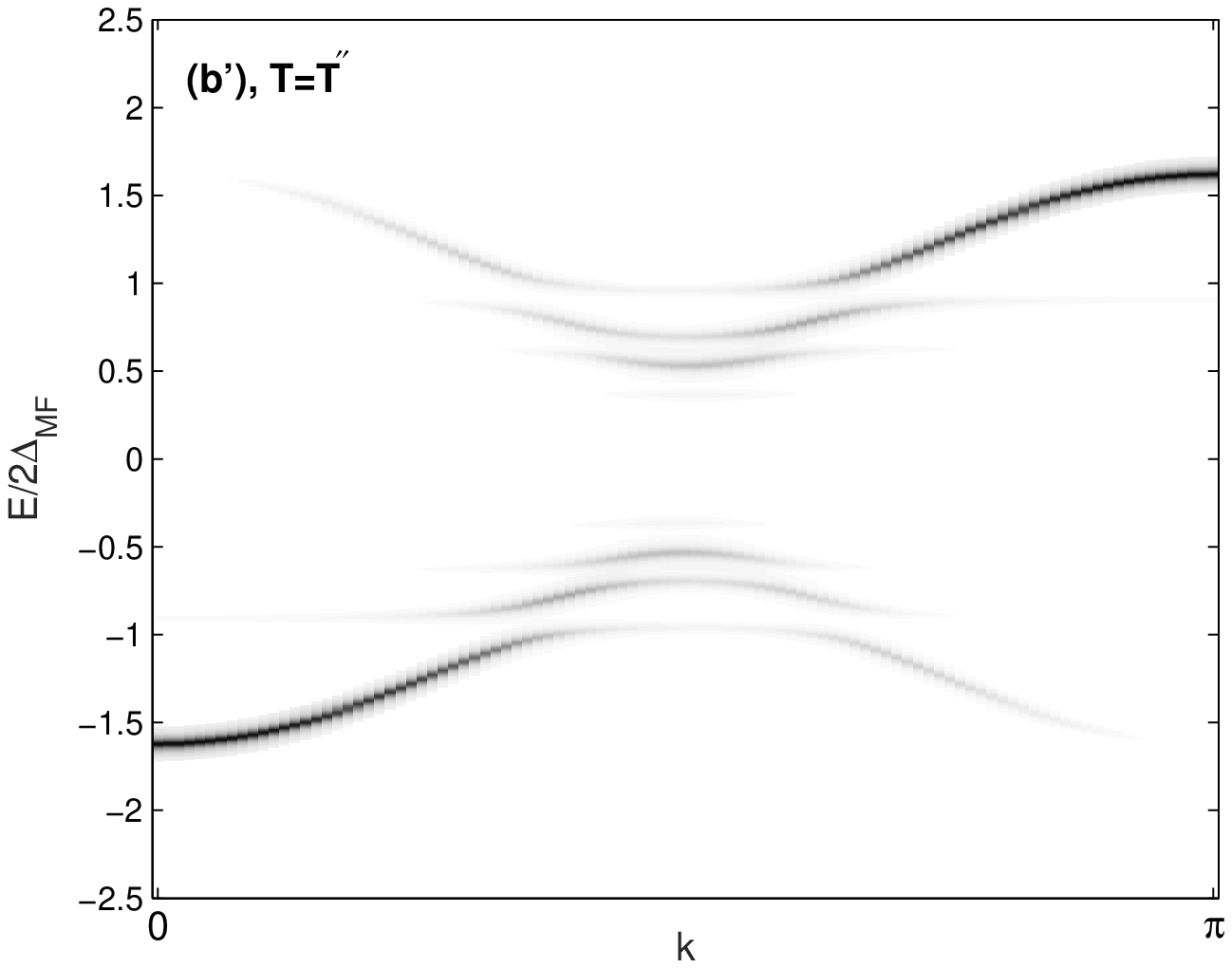}
 \includegraphics[angle=0,width=0.35\columnwidth]{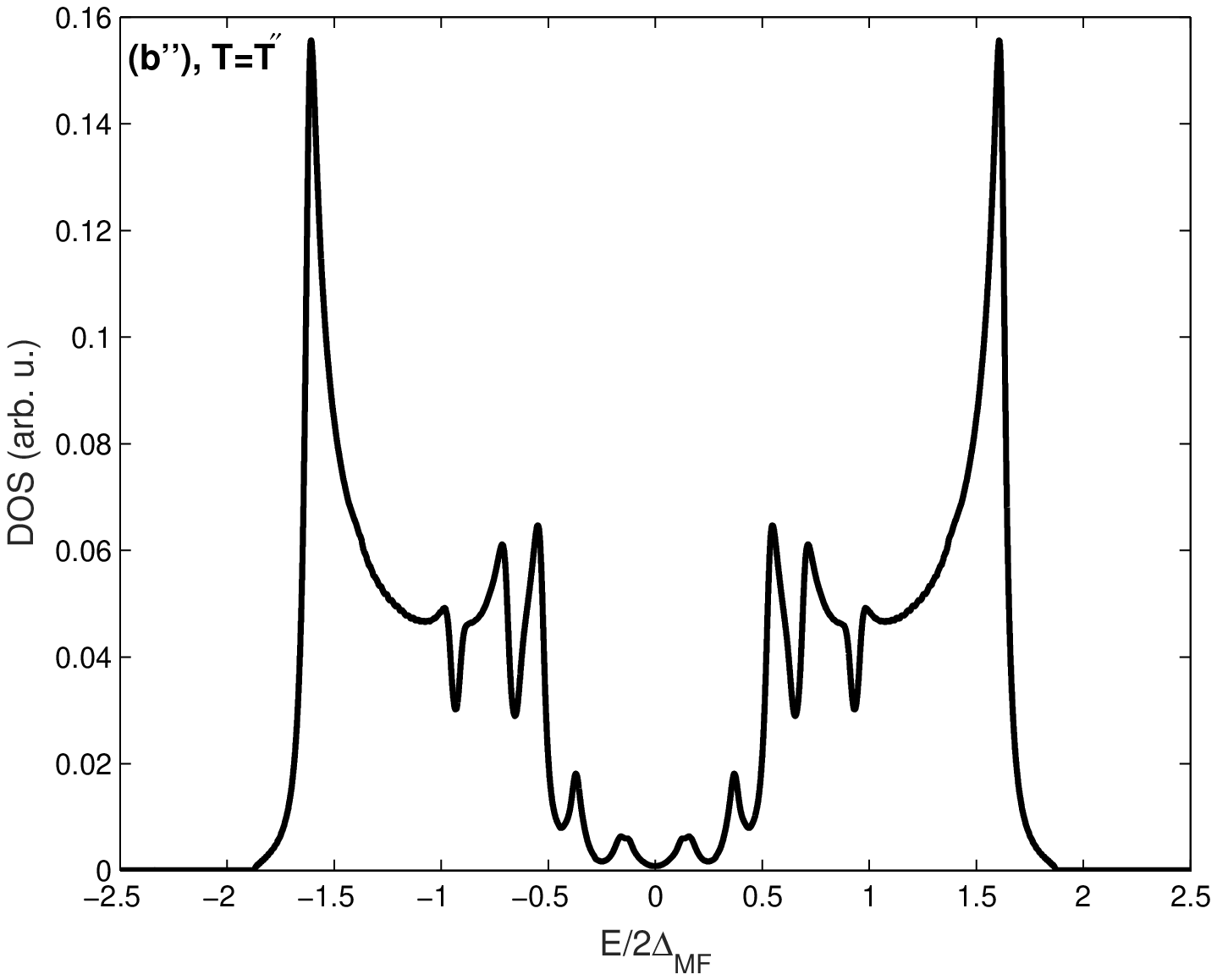}
 \includegraphics[angle=0,width=0.35\columnwidth]{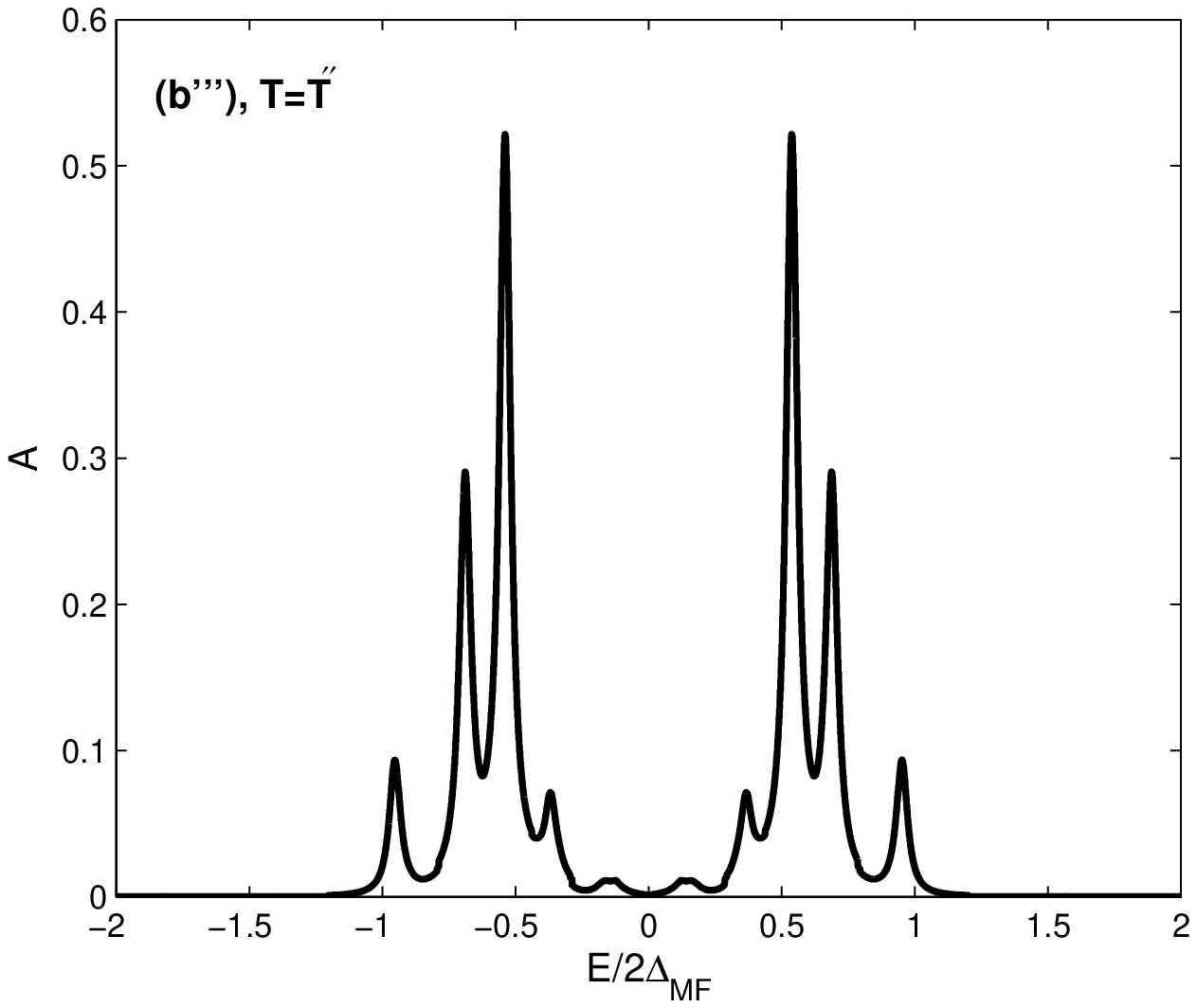}\\
 \includegraphics[angle=0,width=0.35\columnwidth]{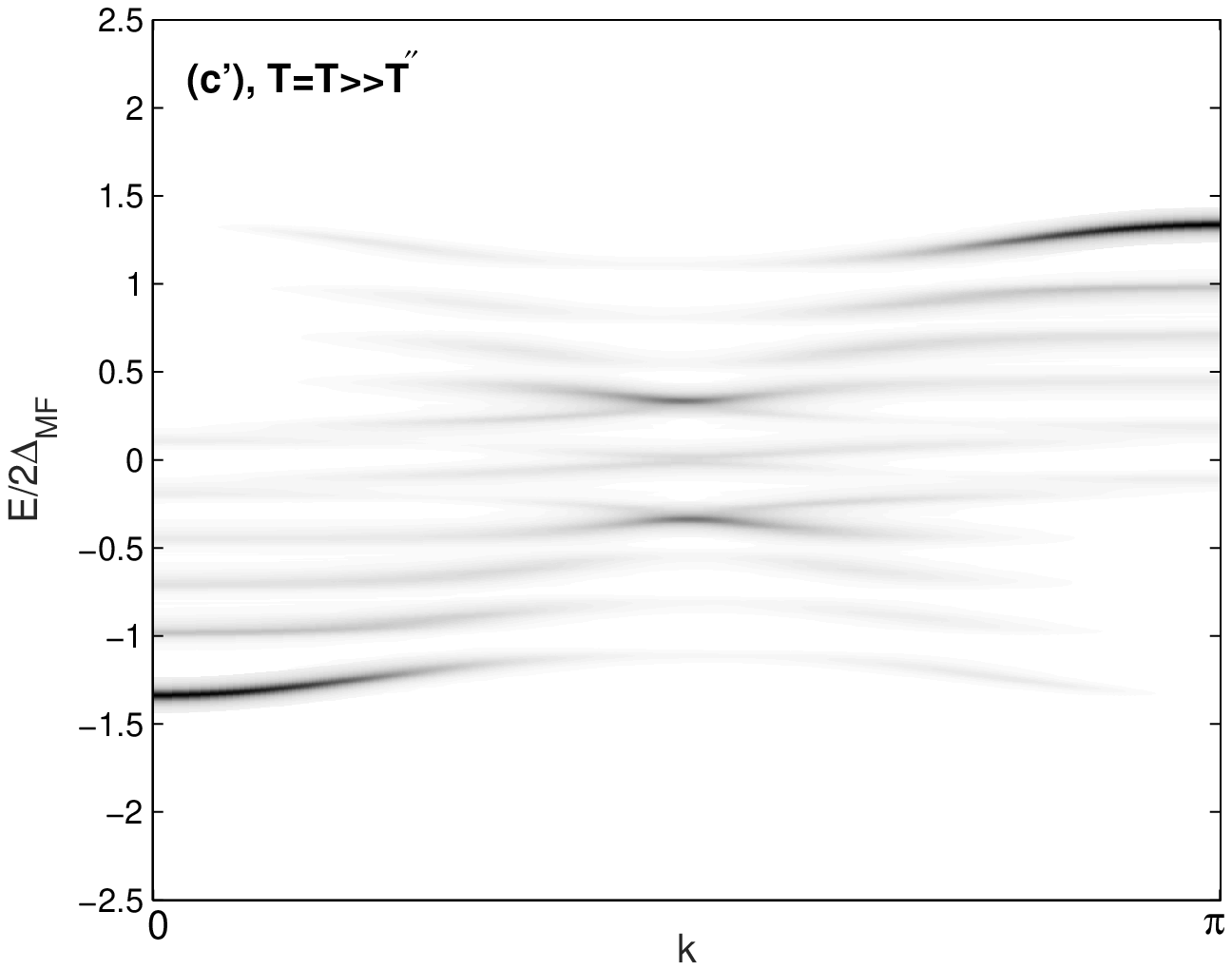}
 \includegraphics[angle=0,width=0.35\columnwidth,height=0.27\columnwidth]{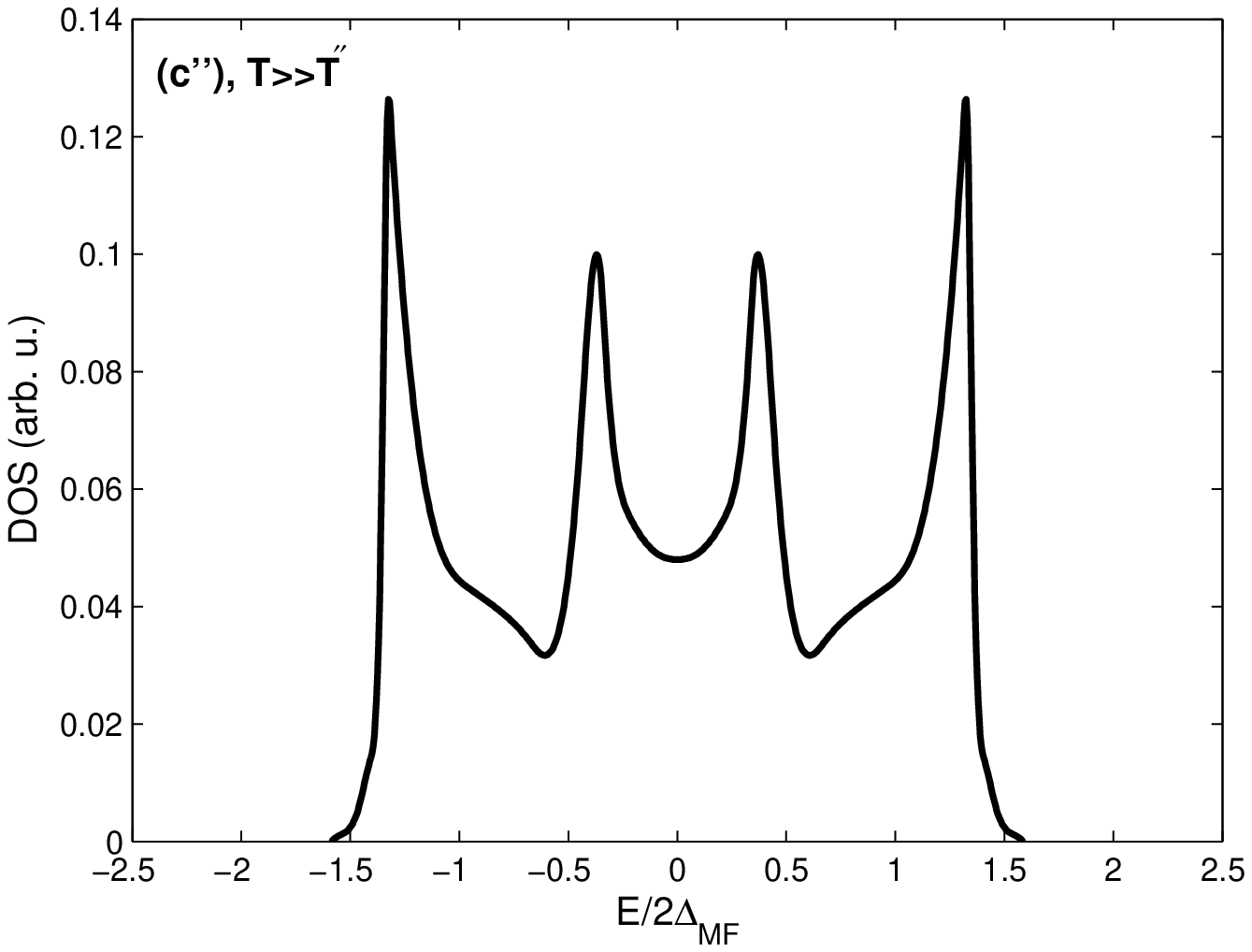}
 \includegraphics[angle=0,width=0.35\columnwidth]{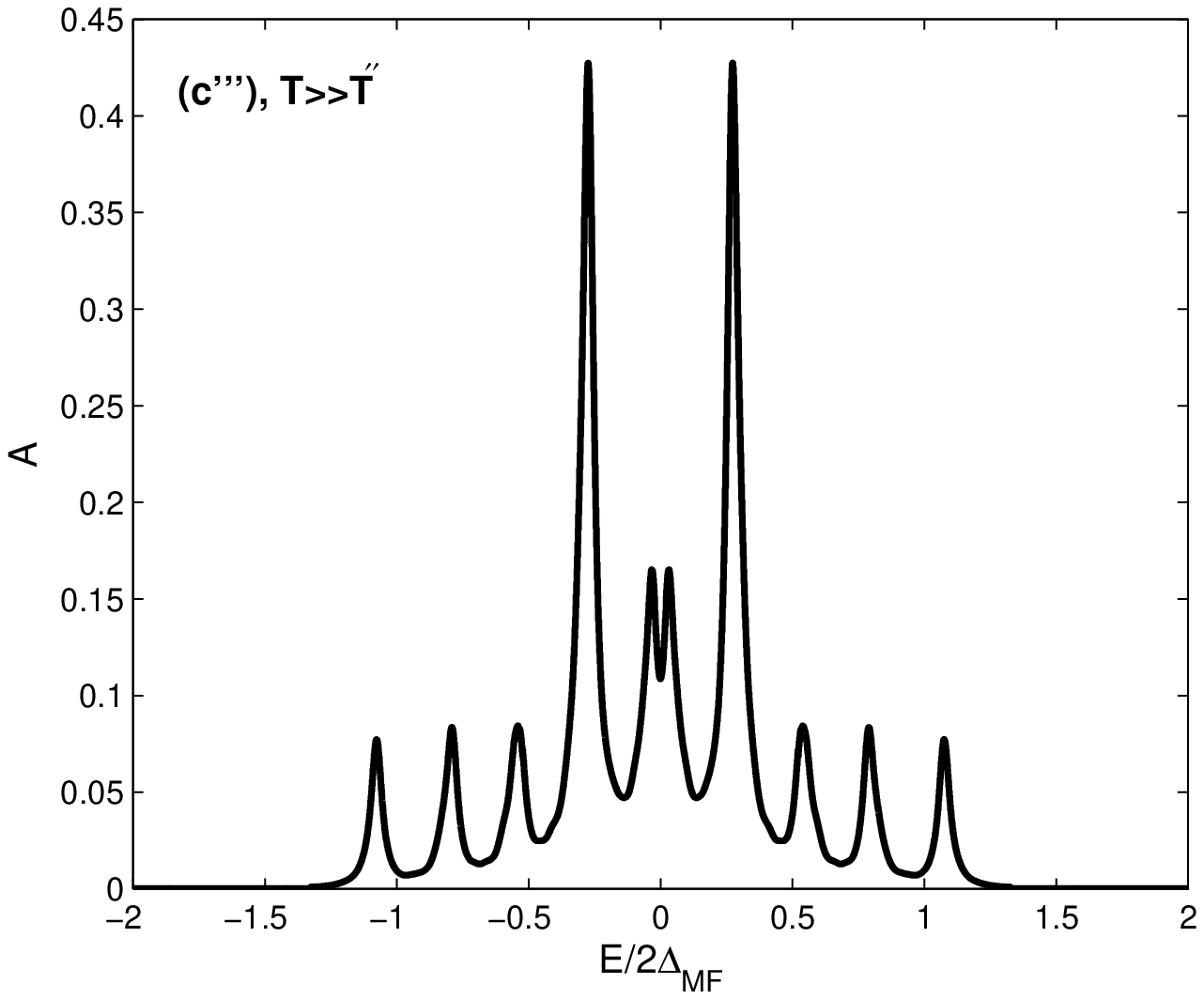}
 \end{array}$
\caption{\label{fig3} Electronic structure (band structure, density of states and the spectral density at the point with the wave vector $k = \pi/2$ at the chemical potential level) in the case of a strong EPI calculated for the three values of temperature (a) $T = 0$, (b) $T = T''$ and (c) $T \gg T''$. All calculations were performed for the following values of parameters: ${\omega _0} = 0.05$eV, $\lambda  = 0.4$, $t = 1$eV, $\delta  = 0.02 - 0.05$eV.}
\end{center}
\end{figure*}
{\sloppy

}
\newpage

\begin{figure*}
\begin{center}
$\begin{array}{cc}
 \includegraphics[angle=0,width=0.50\columnwidth]{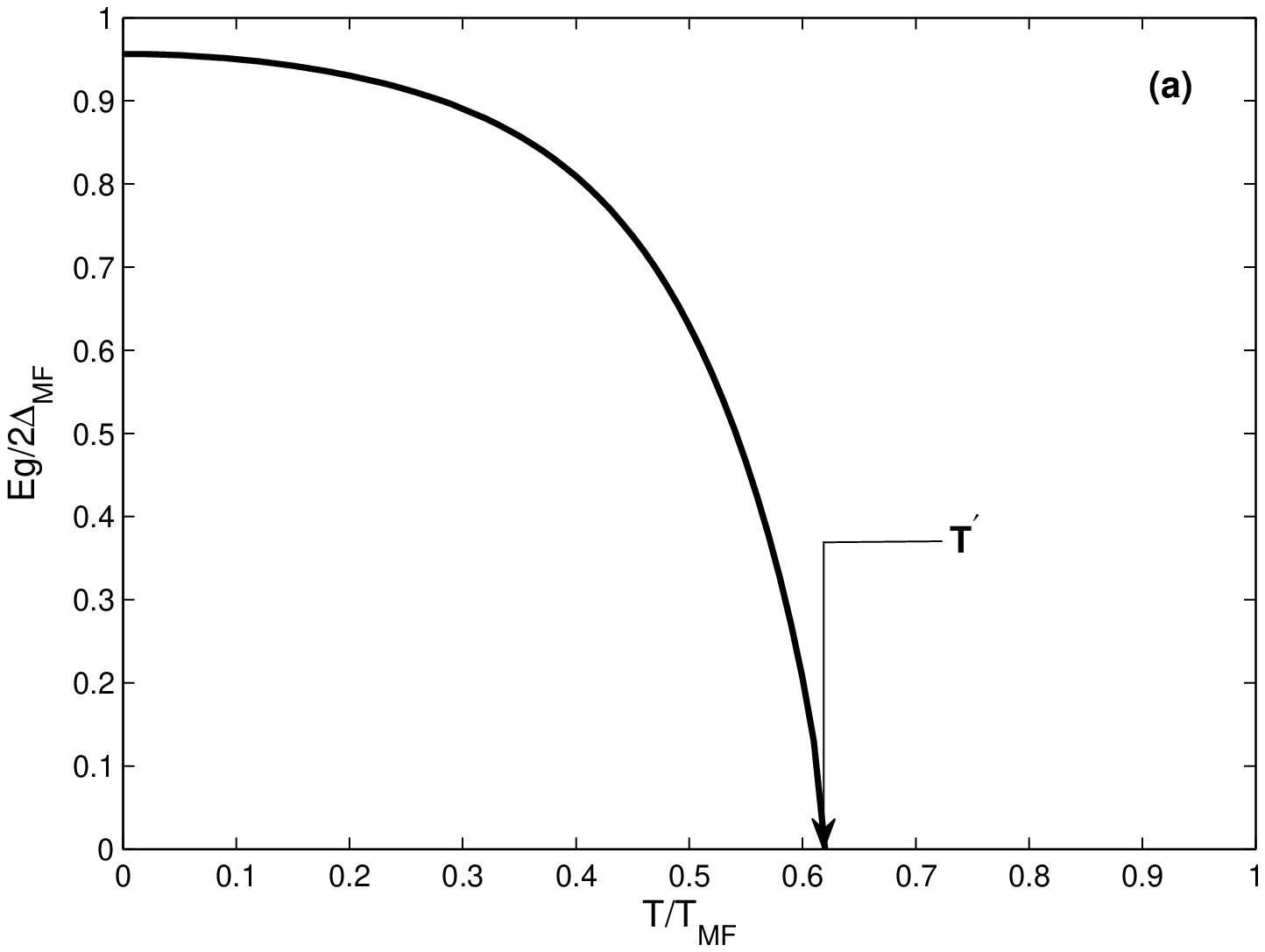}
 \includegraphics[angle=0,width=0.50\columnwidth]{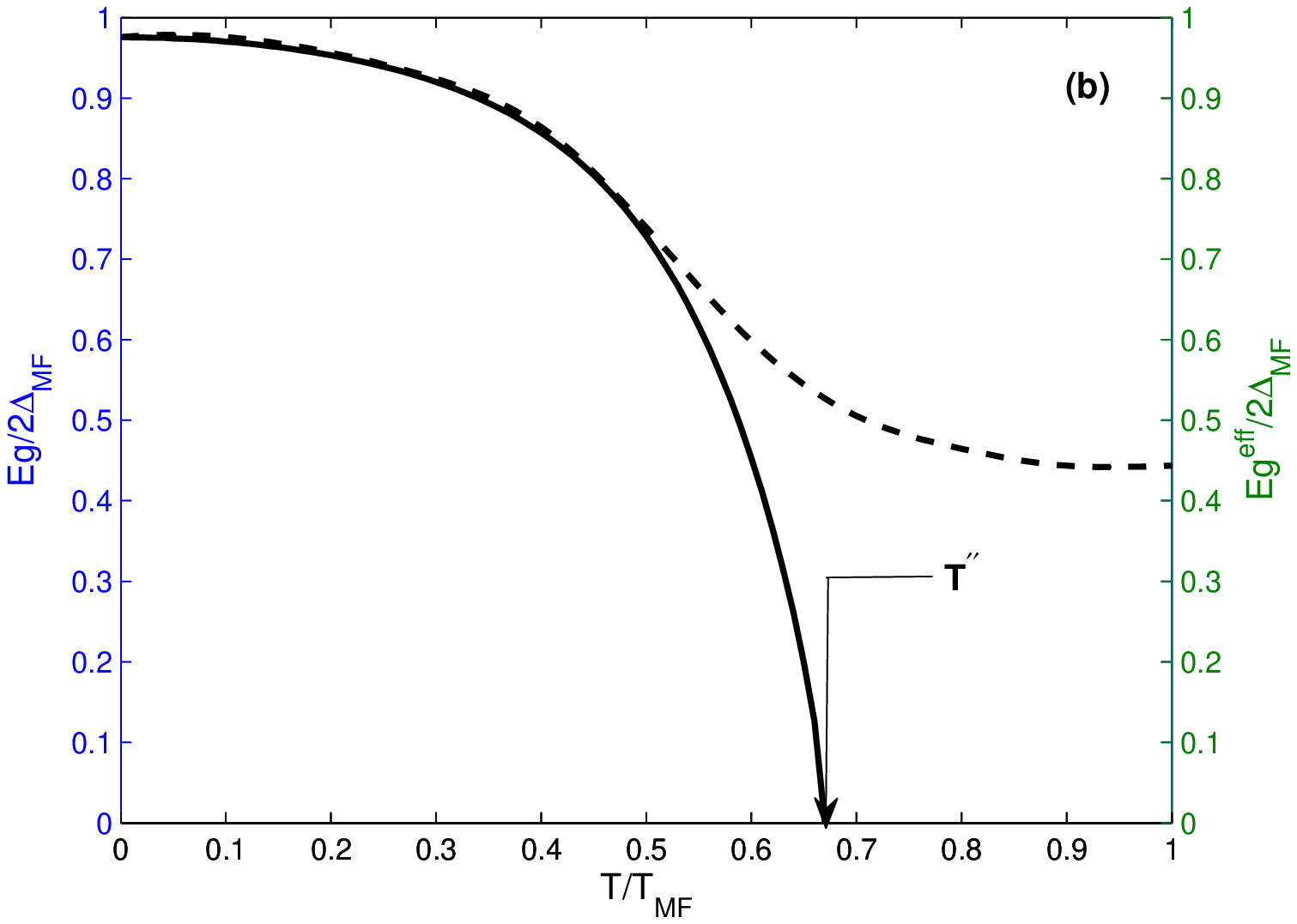}
 \end{array}$
\caption{\label{fig4} Temperature dependence of the dielectric gap ${E_g}$ (solid line) and the effective gap $E_g^{eff}$ (dashed line) in the case of a weak (a) and strong (b) EPI. $2{\Delta _{MF}}$ and ${T_{MF}}$ - the gap and temperature transition value in the mean field theory.}
\end{center}
\end{figure*}
{\sloppy

}

\end{document}